\documentclass[%
 reprint,
 amsmath,amssymb,
 aps,
prb,
]{revtex4-1}

\usepackage{graphicx}
\usepackage{dcolumn}
\usepackage{bm}

\newcommand{\R}{\mathbb{R}}

\newcommand{\ie}{\textit{i.e.}\/, }
\newcommand{\eg}{\textit{e.g.}\/, }
\newcommand{\cf}{\textit{cf.}\/, }

\providecommand*{\mrm}[1]{\mathrm{#1}}
\providecommand*{\unit}[1]{\ensuremath{\mrm{\,#1}}}
\providecommand*{\eu}{\ensuremath{\mrm{e}}}
\providecommand*{\iu}{\ensuremath{\mrm{i}}}

\providecommand*{\diff}{\operatorname{d}\!}
\renewcommand{\Re}{\ensuremath{\mrm{Re}}}	
\renewcommand{\Im}{\ensuremath{\mrm{Im}}}	

\newcommand{\maximize}{\mrm{maximize}}

\newcommand{\subto}{\mrm{subject\ to}}

\def\XXint#1#2#3{{\setbox0=\hbox{$#1{#2#3}{\int}$}
     \vcenter{\hbox{$#2#3$}}\kern-.5\wd0}}

\usepackage[nolist]{acronym}
\usepackage[colorinlistoftodos,prependcaption,textsize=small]{todonotes}
\usepackage[normalem]{ulem}
\usepackage{regexpatch}
\makeatletter
\xpatchcmd{\@todo}{\setkeys{todonotes}{#1}}{\setkeys{todonotes}{inline,#1}}{}{}
\makeatother

\begin{document}


\title{Optical theorems and physical bounds on absorption in lossy media}

\author{Yevhen~Ivanenko}
 \email{yevhen.ivanenko@lnu.se}
\affiliation{%
 Department of Physics and Electrical Engineering, \\ Linn\ae us University, 351 95 V\"{a}xj\"{o}, Sweden.
}%

\author{Mats~Gustafsson}
 \email{mats.gustafsson@eit.lth.se}
\affiliation{%
Department of Electrical and Information Technology, Lund University, Box 118, 221 00 Lund, Sweden.
}%

\author{Sven~Nordebo}
\email{sven.nordebo@lnu.se}
\affiliation{
 Department of Physics and Electrical Engineering, \\ Linn\ae us University, 351 95 V\"{a}xj\"{o}, Sweden.
}%

\date{\today}

\begin{abstract}
Two different versions of an optical theorem for a scattering body embedded inside a lossy background medium are derived in this paper. 
The corresponding fundamental upper bounds on absorption are then obtained in closed form by elementary optimization techniques. 
The first version is formulated in terms of polarization currents (or equivalent currents) inside the scatterer
and generalizes previous results given for a lossless medium. The corresponding bound is referred to here as a variational bound and
is valid for an arbitrary geometry with a given material property. The second version 
is formulated in terms of the T-matrix parameters of 
an arbitrary linear scatterer circumscribed by a spherical volume and gives a new fundamental upper bound on the total absorption of an 
inclusion with an arbitrary material property (including general bianisotropic materials).
The two bounds are fundamentally different as they are based on different assumptions regarding the structure and the material property.
Numerical examples including homogeneous and layered (core-shell) spheres are given to demonstrate that the two bounds 
provide complimentary information in a given scattering problem.
\end{abstract}

\maketitle


\section{Introduction}

Fundamental limits on the scattering and absorption in resonant electromagnetic structures have been considered
in various formulations and applications such as with small dipole scatterers \cite{Tretyakov2014}, antennas \cite{Gustafsson+etal2007a,Sohl+etal2007a},
radar absorbers \cite{Rozanov2000}, high-impedance surfaces \cite{Gustafsson+Sjoberg2011}, 
passive metamaterials \cite{Skaar+Seip2006,Sohl+etal2007c} and optical systems \cite{Miller+etal2016,Shim+etal2019,Molesky+etal2019}. 
Notably, these problems are almost always formulated for a lossless background medium such as vacuum.
In most cases there are very good reasons for doing this, at least when the background losses are sufficiently small. 
In fact, it turns out that the presence of the lossy background medium not only implies an obstructive complication of the analytical derivations, 
it also nullifies the validity of many powerful theorems and constraints, see \eg  \cite{Nordebo+etal2019a,Nordebo+etal2019b,Bohren+Gilra1979,Lebedev+etal1999,Fu+Sun2001,Sudiarta+Chylek2001,Sudiarta+Chylek2002,Yin+Pilon2006,Durant+etal2007a,Mishchenko+etal2017,Mishchenko+Dlugach2019} with references. 
In essence, the problem is that any optical theorem for a scattering body embedded in a lossy medium will depend on the geometry of the scatterer.
A vivid illustration of this is the optical theorem and the associated upper bounds on 
dipole scattering and absorption of a small (dipole) scatterer in a lossless medium which are based solely on the polarizability of the scatterer \cite{Tretyakov2014},
and which no longer is valid for a lossy background \cite{Yin+Pilon2006,Mishchenko+etal2017,Nordebo+etal2019b}.
Hence, for a lossy background medium the associated optical theorems must be modified, and
it can be expected that any analytical results regarding the optimal absorption will become dependent on the geometry of the scatterer,
see \eg \cite{Nordebo+etal2019a,Nordebo+etal2019b}.

There is a number of application areas where the surrounding losses clearly cannot be neglected. This includes typically medical applications
such as localized electrophoretic heating of a bio-targeted and electrically charged gold nanoparticle suspension as a radiotherapeutic hyperthermia-based method to treat cancer, 
\cf \cite{Corr+etal2012,Sassaroli+etal2012,Collins+etal2014,Nordebo+etal2017a,Dalarsson+etal2017a}. Corresponding applications in the optical domain
are concerned with light in biological tissue \cite{Duck1990}, and the use of gold nanoparticles for plasmonic photothermal therapy \cite{Huang+etal2008}.
Surface-enhanced biological sensing with molecular monolayer spectroscopy is another related application, see \eg \cite{Maier2007}. 
In photonic applications and in plasmonics, dielectric substrates based on polymeric media are usually considered to be lossless at optical frequencies \cite{Progelhof+etal1971,Krasnok+etal2012}. 
However, some of the substances that are used can also show significant losses such as with z-doped PMMA materials \cite{AlTaay+etal2015}.
Another important medical application is concerned with implantable antennas that are used in telemetry applications as a part of communication link of health-monitoring and health-care systems \cite{Skrivervik+etal2019,Merli+etal2012}. 
The presence of losses in the background medium affects the reliability of such links, especially the performance of in-body antennas. In this application, however, the aim is to {\em reduce} the amount of power absorbed by a human body, and this can be achieved by encapsulation of the implant by a biocompatible insulator \cite{Merli+etal2010}. Finally, we mention the terrestrial gaseous atmosphere which has a diversity of rotational-vibrational absorption bands ranging from microwave to optical 
frequencies \cite{Liou2002,Gordon+etal2017}.
Important applications include antennas and short range communications at 60 \unit{GHz} \cite{Park+Rappaport2007,Hawkins+etal1985,Wang+etal2012,Vleck1947,Meeks+Lilley1963} (absorption bands of oxygen)
as well as the study of radiative transfer in the presence of aerosols and cloud particles in the atmosphere \cite{Liou2002,Mishchenko+Dlugach2019}.

In this paper we present two different versions of an optical theorem and the associated absorption bounds for a scattering body embedded inside a lossy background medium. 
These versions are based on the interior and the exterior fields, \ie the equivalent (polarization) currents inside the scatterer and the T-matrix parameters of the scatterer, respectively. 
The two versions of the optical theorem express the same power balance, and
yet they are fundamentally different (and hence complimentary), since they are based on different assumptions regarding the properties of the scatterer.
The first version is an extension to lossy media regarding the absorption bounds given in \cite{Miller+etal2016}, and is valid for an arbitrary geometry with a given material property.
Even though the basic optimization technique is 
the same as in \cite{Miller+etal2016}, it is demonstrated how the seemingly trivial extension to lossy media 
insidiously requires a careful analysis where it is not sufficient just to replace a real-valued wave number for a complex-valued one. 
In particular, the new optical theorem shows that the extinct power must be expressed as an affine form in the equivalent currents implying subtle changes 
in the final form of the fundamental bound on absorption.
The second version is a refinement of the fundamental bounds on multipole absorption given in \cite{Nordebo+etal2019a}, 
and is valid for a spherical geometry with an arbitrary material property.
In particular, we formulate an optical theorem and derive the associated absorption bound for the total fields including all the electric and magnetic multipoles. It is proved that the bound is valid not only for a rotationally invariant sphere as in \cite{Nordebo+etal2019a}, but also for general heterogeneous bianisotropic materials.
We prove also that the new bound, which is given by a multipole summation formula, is convergent whenever there are non-zero losses
in the exterior domain. In this way, the results also provide a new way to determine the number of useful multipoles in a given scattering problem, 
\ie as a function of the electrical size of the scatterer as well as of the losses in the exterior domain.
Through the numerical examples, we show that the new fundamental bounds give complementary information on the absorption of scattering objects in lossy media. The derived bounds are applicable for 
arbitrary objects made of arbitrary materials, which gives a possibility to find such an electrically small structure that has an absorption peak close to the fundamental bounds.

The rest of the paper is organized as follows: In Section \ref{sect:opttheintfields} is given the optical theorem based on the interior fields and 
in Section \ref{sect:varbound} the corresponding bounds on absorption by variational calculus.
In Section \ref{sect:multbound}, we consider the optical theorem and the associated bounds based on the exterior fields using the T-matrix formalism.
In Section \ref{sect:numexamples} is illustrated the numerical examples, and the paper is summarized in Section \ref{sect:summaryandconclusions}.
Finally, in Appendix~\ref{sect:varcalculus} is shown the derivation of the maximal absorbed power based on calculus of variations, and in Appendix \ref{sect:spherical} is put the most important definitions and formulas that are used regarding the spherical vector wave expansion.

\section{Optical theorem based on the interior fields}\label{sect:opttheintfields}
\subsection{Notation and conventions}
The electric and magnetic field intensities $\bm{E}$ and $\bm{H}$
are given in SI-units \cite{Jackson1999} and the time convention for time harmonic fields (phasors) is given by $\eu^{-\iu\omega t}$,
where $\omega$ is the angular frequency and $t$ the time. 
Let $\mu_0$, $\epsilon_0$, $\eta_0$ and $\mrm{c}_0$ denote the permeability, the permittivity, the wave impedance and
the speed of light in vacuum, respectively, and where $\eta_0=\sqrt{\mu_0/\epsilon_0}$ and $\mrm{c}_0=1/\sqrt{\mu_0\epsilon_0}$.
The wave number of vacuum is given by $k_0=\omega\sqrt{\mu_0\epsilon_0}$, and hence $\omega\mu_0=k_0\eta_0$ and $\omega\epsilon_0=k_0\eta_0^{-1}$. 
The real and imaginary parts and the complex conjugate of a complex number $\zeta$ are denoted by $\Re\left\{\zeta\right\}$, $\Im\left\{\zeta\right\}$ and $\zeta^*$, respectively. For dyadics, the notation $(\cdot)^\dagger$ denotes the Hermitian transpose.

\subsection{Extinction, scattering and absorption}
\begin{figure}[htb]
\begin{center}
\includegraphics[width=0.25\textwidth]{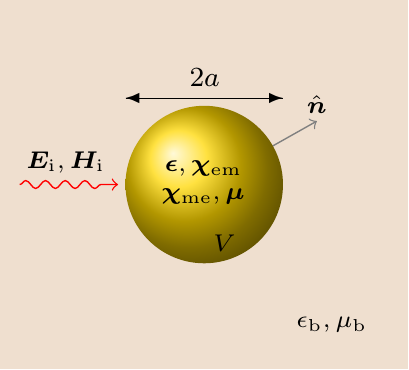}
\end{center}
\vspace{-5mm}
\caption{Problem setup. Here, $\epsilon_\mrm{b}$ and $\mu_\mrm{b}$ denote the relative permittivity and permeability of the passive background medium, respectively, and $\hat{\bm{n}}$ the outward unit vector.
}
\label{fig:matfig11}
\end{figure}

Consider a scattering problem consisting of a scattering body $V$ bounded by the surface $\partial V$ and which is embedded in an infinite homogeneous and isotropic  
background medium having relative permeability $\mu_\mrm{b}$ and relative permittivity $\epsilon_\mrm{b}$, see Fig.~\ref{fig:matfig11}.
The scatterer $V$ consists of a 
linear material bounded by a finite open set with volume denoted by the same letter $V$.
The background medium also consists of a passive material and hence $\Im\{\mu_\mrm{b}\}\geq 0$ and $\Im\{\epsilon_\mrm{b}\}\geq 0$.
The  incident (i) and scattered (s) fields satisfy the following Maxwell's equations with respect to the background medium
\begin{equation}\label{eq:MaxwellEiEs}
\left\{\begin{array}{l}
\nabla\times\bm{E}_\mrm{\{i,s\}}=\iu k_0\eta_0\mu_\mrm{b}\bm{H}_\mrm{\{i,s\}}, \vspace{0.2cm} \\
\nabla\times\bm{H}_\mrm{\{i,s\}}=-\iu k_0\eta_0^{-1}\epsilon_\mrm{b}\bm{E}_\mrm{\{i,s\}},
\end{array}\right.
\end{equation}
in the exterior region $\R^3\setminus V$ and the total fields are denoted $\bm{E}=\bm{E}_\mrm{i}+\bm{E}_\mrm{s}$ and $\bm{H}=\bm{H}_\mrm{i}+\bm{H}_\mrm{s}$.
It is noted that the incident fields satisfy \eqref{eq:MaxwellEiEs} in the whole of $\R^3$.

The interior scattering medium is characterized by the following constitutive relations for a general bianisotropic linear material
\begin{equation}\label{eq:constitutive}
\left\{\begin{array}{l}
\bm{D}=\epsilon_0\bm{\epsilon}\cdot\bm{E}+\frac{1}{\mrm{c}_0}\bm{\chi}_\mrm{em}\cdot\bm{H}, \vspace{0.2cm} \\
\bm{B}=\frac{1}{\mrm{c}_0}\bm{\chi}_\mrm{me}\cdot\bm{E} +\mu_0\bm{\mu}\cdot\bm{H},
\end{array}\right.
\end{equation}
where $\bm{B}$ is the magnetic flux density and $\bm{D}$ the electric flux density and where the relative permittivity and permeability dyadics are
$\bm{\epsilon}=\epsilon_\mrm{b}\bm{I}+\bm{\chi}_\mrm{ee}$ and $\bm{\mu}=\mu_\mrm{b}\bm{I}+\bm{\chi}_\mrm{mm}$, respectively, and where $\bm{\chi}_\mrm{ee}$, $\bm{\chi}_\mrm{mm}$,
$\bm{\chi}_\mrm{em}$ and $\bm{\chi}_\mrm{me}$ are dimensionless susceptibility dyadics.
By following the standard volume equivalence principles \cite{Volakis+Sertel2012}, the Maxwell's equations for the interior region $V$ 
\begin{equation}\label{eq:generalMaxwell}
\left\{\begin{array}{l}
\nabla\times\bm{E}=\iu k_0\bm{\chi}_\mrm{me}\cdot\bm{E} +\iu k_0\eta_0\bm{\mu}\cdot\bm{H}, \vspace{0.2cm} \\
\nabla\times\bm{H}=-\iu k_0\eta_0^{-1}\bm{\epsilon}\cdot\bm{E}-\iu k_0\bm{\chi}_\mrm{em}\cdot\bm{H},
\end{array}\right.
\end{equation}
can now be reformulated in terms of the background medium as
\begin{equation}\label{eq:Maxwelleqsources}
\left\{\begin{array}{l}
\nabla\times\bm{E}=\iu k_0\eta_0\mu_\mrm{b}\bm{H}-\bm{J}_\mrm{m}, \vspace{0.2cm} \\
\nabla\times\bm{H}=-\iu k_0\eta_0^{-1}\epsilon_\mrm{b}\bm{E}+\bm{J}_\mrm{e},
\end{array}\right.
\end{equation}
which is just \eqref{eq:generalMaxwell} rewritten based on the equivalent electric and magnetic contrast currents 
\begin{equation}\label{eq:defeqsources}
\left\{\begin{array}{l}
\bm{J}_\mrm{e}=-\iu k_0\eta_0^{-1}\bm{\chi}_\mrm{ee}\cdot\bm{E}-\iu k_0\bm{\chi}_\mrm{em}\cdot\bm{H}, \vspace{0.2cm} \\
\bm{J}_\mrm{m}=-\iu k_0\bm{\chi}_\mrm{me}\cdot\bm{E}-\iu k_0\eta_0\bm{\chi}_\mrm{mm}\cdot\bm{H}.
\end{array}\right.
\end{equation}

The power balance at the surface $\partial V$ just outside  $V$ (where $\bm{E}=\bm{E}_\mrm{i}+\bm{E}_\mrm{s}$) 
is obtained by using the corresponding Poynting's vectors and can be expressed as
\begin{equation}\label{eq:Opticaltheorem}
P_\mrm{a}=-P_\mrm{s}+P_\mrm{t}+P_\mrm{i},
\end{equation}
where $P_\mrm{a}$, $P_\mrm{s}$, $P_\mrm{t}$ and $P_\mrm{i}$ are the absorbed, scattered, extinct (total) and the incident powers, respectively,
defined by
\begin{eqnarray}
P_\mrm{a} & = & -\frac{1}{2}\Re\left\{ \int_{\partial V}\bm{E}\times\bm{H}^* \cdot \hat{\bm{n}}\diff S \right\},  \label{eq:Padef} \\
P_\mrm{s} & = & \frac{1}{2}\Re\left\{ \int_{\partial V}\bm{E}_\mrm{s}\times\bm{H}_\mrm{s}^* \cdot \hat{\bm{n}}\diff S \right\}, \label{eq:Psdef}\\
P_\mrm{t} & = & -\frac{1}{2}\Re\left\{ \int_{\partial V}\left(\bm{E}_\mrm{i}\times\bm{H}_\mrm{s}^* + \bm{E}_\mrm{s}\times\bm{H}_\mrm{i}^*\right)  \cdot \hat{\bm{n}}\diff S \right\}, \label{eq:Ptdef} \\
P_\mrm{i} & = & -\frac{1}{2}\Re\left\{ \int_{\partial V}\bm{E}_\mrm{i}\times\bm{H}_\mrm{i}^* \cdot \hat{\bm{n}}\diff S \right\}, \label{eq:Pidef}
\end{eqnarray}
and where the surface integrals are defined with an outward unit normal $\hat{\bm{n}}$, 
see also \cite[Eq.~(3.19)]{Bohren+Huffman1983}. 
Based on the Poyntings theorem (the divergence theorem) $\int_{\partial V}\bm{E}\times\bm{H}^* \cdot  \hat{\bm{n}}\diff S=\int_{V} \left(\bm{H}^*\cdot\nabla\times\bm{E}-\bm{E}\cdot\nabla\times\bm{H}^*\right)  \diff v$,
and by employing the following identities on $\partial V$ 
\begin{equation}\label{eq:boundaryconditions}
\left\{\begin{array}{l}
\hat{\bm{n}}\times(\bm{E}_\mrm{i}+\bm{E}_\mrm{s})=\hat{\bm{n}}\times\bm{E}, \vspace{0.2cm} \\
\hat{\bm{n}}\times(\bm{H}_\mrm{i}+\bm{H}_\mrm{s})=\hat{\bm{n}}\times\bm{H},
\end{array}\right.
\end{equation}
the vector identity $\hat{\bm{n}}\cdot\bm{X}\times\bm{Y}=\hat{\bm{n}}\times\bm{X}\cdot\bm{Y}$ and the Maxwell's equations \eqref{eq:generalMaxwell}, it is possible to show that \eqref{eq:Opticaltheorem} gives an optical theorem for the lossy background where
\begin{eqnarray}
P_\mrm{a} & = & \frac{k_0}{2\eta_0}\Im\left\{\int_{V}\bm{F}^*\cdot \bm{M}_\mrm{a}\cdot\bm{F}\diff v\right\},  \label{eq:Paexpr}\\
P_\mrm{t} & = & \frac{k_0}{2\eta_0}\Im\left\{\int_{V}\bm{F}_\mrm{i}^*\cdot \bm{M}_\mrm{t}\cdot\bm{F}\diff v\right\} -2P_\mrm{i}, \label{eq:Ptexpr} \\
P_\mrm{i} & = & \frac{k_0}{2\eta_0}\Im\left\{\int_{V}\bm{F}_\mrm{i}^*\cdot \bm{M}_\mrm{b}\cdot\bm{F}_\mrm{i}\diff v\right\}, \label{eq:Piexpr}
\end{eqnarray}
and which are based solely on the interior fields. 
Here, the field quantities are defined as 
\begin{equation}\label{eq:FandFidef}
\bm{F}=\left(\begin{array}{c}
\bm{E}  \\
\eta_0\bm{H}
\end{array}\right), \quad 
\bm{F}_\mrm{i}=\left(\begin{array}{c}
\bm{E}_\mrm{i}  \\
\eta_0\bm{H}_\mrm{i}
\end{array}\right),
\end{equation}
and the material dyadics are given by 
\begin{equation}\label{eq:Madef}
\bm{M}_\mrm{a}=\left(\begin{array}{cc}
\bm{\epsilon} & \bm{\chi}_\mrm{em} \\
\bm{\chi}_\mrm{me} & \bm{\mu}
\end{array}\right)
=\bm{\chi}+\bm{M}_\mrm{b}
\end{equation}
where
\begin{equation}\label{eq:chiMbdef}
\bm{\chi}=\left(\begin{array}{cc}
\bm{\chi}_\mrm{ee} & \bm{\chi}_\mrm{em} \\
\bm{\chi}_\mrm{me} & \bm{\chi}_\mrm{mm}
\end{array}\right), \quad 
\bm{M}_\mrm{b}=\left(\begin{array}{cc}
\epsilon_\mrm{b}\bm{I} & \bm{0} \\
\bm{0} & \mu_\mrm{b}\bm{I}
\end{array}\right),
\end{equation}
and
\begin{equation}\label{eq:Mtdef}
\bm{M}_\mrm{t}=
\left(\begin{array}{cc}
\bm{\epsilon}-\epsilon_\mrm{b}^*\bm{I} & \bm{\chi}_\mrm{em} \\
\bm{\chi}_\mrm{me} & \bm{\mu}-\mu_\mrm{b}^*\bm{I}
\end{array}\right)
=\bm{\chi}+\iu 2\Im\{\bm{M}_\mrm{b}\}.
\end{equation}
It is noted that \eqref{eq:Paexpr} through \eqref{eq:Mtdef}  generalizes previous expressions which have been given 
for a lossless exterior medium \cite[Eqs.~(4) through (7) on p.~3338]{Miller+etal2016} where $P_\mrm{i}=0$, and both $\bm{M}_\mrm{a}$ 
and  $\bm{M}_\mrm{t}$ given above can be replaced by the susceptibility dyadic $\bm{\chi}$.

It is observed that $P_\mrm{a}$ is represented by a positive definite (strictly convex) quadratic form and  $P_\mrm{t}$ by an affine form in the field quantities.
Note in particular the additional power balancing term $-2P_\mrm{i}$ that is present in \eqref{eq:Ptexpr}.
Finally, it is noted that in the present formulation it is sufficient to derive three terms as in \eqref{eq:Paexpr} through \eqref{eq:Piexpr}
since the fourth term $P_\mrm{s}$ will then be given by the optical theorem \eqref{eq:Opticaltheorem}.

\section{Fundamental bounds on absorption by variational calculus}\label{sect:varbound}
The fundamental bounds on absorption derived in \cite{Miller+etal2016} are generalized below
for the case with a lossy background medium. The derivation is based on the 
optical theorem expressed in \eqref{eq:Opticaltheorem} 
together with \eqref{eq:Paexpr} through \eqref{eq:Mtdef} above.

\subsection{General bianisotropic media}
The optimization problem of interest is given by
\begin{eqnarray}\label{eq:Paoptdef}
\begin{array}{llll}
	& \maximize & & P_\mrm{a}  \\    
	& \subto & &  P_\mrm{s} \geq 0,
\end{array}
\end{eqnarray}
where the optimization is with respect to the interior fields $\bm{F}$ of the structure, and where the scattered power
$P_\mrm{s}=-P_\mrm{a}+P_\mrm{t}+P_\mrm{i}$ is used as the non-negative constraint. 
In this case, the constraint is organized as $-P_\mrm{a}+P_\mrm{t}+2P_\mrm{i}-P_\mrm{i}\geq 0$
where $P_\mrm{a}$ is the positive definite quadratic form expressed in \eqref{eq:Paexpr}, $P_\mrm{t}+2P_\mrm{i}$ is the linear form given by \eqref{eq:Ptexpr}
and $P_\mrm{i}$ is given by \eqref{eq:Piexpr}. This is a convex maximization problem having
a unique solution at the boundary of the feasible region (active constraint), \cf \cite{Luenberger1969}.

By using the method of Lagrange multipliers~\cite{Adams1995} and variational calculus, it can be shown that the optimal bound on absorbed power $P_\mrm{a}^\mrm{opt}$ is given by
\begin{equation}\label{eq:Paopt}
P_\mrm{a}^\mrm{opt}=\frac{k_0\alpha^2}{8\eta_0}\int_{V}
\bm{F}_\mrm{i}^*\cdot\bm{M}_\mrm{t}\cdot\left(\Im\{\bm{M}_\mrm{a}\} \right)^{-1}\cdot\bm{M}_\mrm{t}^\dagger\cdot\bm{F}_\mrm{i}\diff v,
\end{equation}
see the detailed derivation of the result \eqref{eq:PaoptApp_res} in Appendix~\ref{sect:maxPabs}.
The parameter $\alpha$ is found by inserting the stationary solution \eqref{eq:stationarysol} into the active constraint in \eqref{eq:Paoptdef}, yielding the quadratic equation
\begin{equation}\label{eq:alpharoots}
\alpha^2+2\alpha=q,
\end{equation}
where
\begin{equation}\label{eq:qdef}
q=\frac{\displaystyle 4\int_{V}\bm{F}_\mrm{i}^*\cdot\Im\left\{\bm{M}_\mrm{b}\right\}\cdot\bm{F}_\mrm{i}\diff v}
{\displaystyle\int_{V}\bm{F}_\mrm{i}^*\cdot\bm{M}_\mrm{t}\cdot\left(\Im\{\bm{M}_\mrm{a}\} \right)^{-1}\cdot\bm{M}_\mrm{t}^\dagger\cdot\bm{F}_\mrm{i}\diff v}.
\end{equation}
The denominator in \eqref{eq:qdef} is convex in $\bm{M}_\mrm{a}$ for $\Im\{\bm{M}_\mrm{a}\}>0$, and thus by its minimization, it can be shown that \eqref{eq:qdef} is maximal for $\bm{M}_\mrm{a}=\bm{M}_\mrm{b}$ implying that $q\leq1$, see the proof in Appendix~\ref{sect:minq}.
The maximizing root of \eqref{eq:alpharoots} is hence given by
\begin{equation}\label{eq:alphadef}
\alpha=-1-\sqrt{1-q}.
\end{equation}

The expression \eqref{eq:Paopt} together with \eqref{eq:Madef} through \eqref{eq:Mtdef}, \eqref{eq:qdef} and \eqref{eq:alphadef} generalizes the previous result in \cite{Miller+etal2016} which has been given
for a lossless exterior medium. In particular, by considering a lossless exterior medium with \eg $\mu_\mrm{b}=\epsilon_\mrm{b}=1$,
it is seen that $q=0$ (which implies that $0\leq q\leq 1$), $\alpha=-2$, $\lambda=2$, $\bm{M}_\mrm{t}=\bm{\chi}$ and $\Im\{\bm{M}_\mrm{a}\}=\Im\{\bm{\chi}\}$, 
so that \eqref{eq:Paopt} reproduces the corresponding result in \cite[Eq.~(23b) on p.~3342]{Miller+etal2016}. 
Optimization of the scattered power can be treated similarly.

\subsection{Piecewise homogeneous and isotropic dielectric structures}
Important special cases are with the optimal absorption of piecewise homogeneous and isotropic dielectric structures in a lossy surrounding dielectric medium. 
In this case, the problem only involves the electric losses and we can simplify the notation by writing 
$\bm{F}=\bm{E}$, $\bm{F}_\mrm{i}=\bm{E}_\mrm{i}$ and $\bm{\chi}_i=\bm{\chi}_{\mrm{ee},i}=(\epsilon_i-\epsilon_\mrm{b})\bm{I}$, where $i=1,\ldots,N$ is related to the corresponding homogeneous component of the composed scatterer. Note that for $N=1$, the problem simplifies to the homogeneous structure.
The associated material dyadics are given by
\begin{equation}\label{eq:dieletricdyadics}
\bm{M}_{\mrm{a},i}=\epsilon_i\bm{I}, \quad \bm{M}_\mrm{b}=\epsilon_\mrm{b}\bm{I}, \quad  \bm{M}_{\mrm{t},i}=(\epsilon_i-\epsilon_\mrm{b}^*)\bm{I},
\end{equation}
and the expression \eqref{eq:Paopt} becomes
\begin{equation}\label{eq:Paopt2}
P_\mrm{a}^\mrm{var}=\frac{k_0\alpha^2}{8\eta_0}\sum_{i=1}^N\frac{\left|\epsilon_i-\epsilon_\mrm{b}^*\right|^2}{\Im\{\epsilon_i\}}\int_{V_i}\left|\bm{E}_\mrm{i}(\bm r)\right|^2\diff v,
\end{equation}
for the total volume of scattering body $V=\sum_i V_i$, $i=1,\ldots,N$, and where $\alpha$ is given by \eqref{eq:alphadef}, and $q$ is obtained from \eqref{eq:qdef} as
\begin{equation}\label{eq:qdef2}
q=\displaystyle\cfrac{4\Im\{\epsilon_\mrm{b}\}\displaystyle\int_{V}\left|\bm{E}_\mrm{i}(\bm r)\right|^2\diff v}{\displaystyle\sum_{i=1}^N\cfrac{\left|\epsilon_i-\epsilon_\mrm{b}^*\right|^2}{\Im\{\epsilon_i\}}\displaystyle\int_{V_i}\left|\bm{E}_\mrm{i}(\bm r)\right|^2\diff v}.
\end{equation}
Assume now that the scatterer is an $N$-layered sphere $V_a$ of total radius $a$, $N\geq1$, and the incident field is a plane wave $\bm{E}_\mrm{i}(\bm r)=\bm{E}_0\eu^{\mrm{i} k_\mrm{b}\hat{\bm{k}}\cdot\bm{r}}$
with vector amplitude $\bm{E}_0$, propagation direction $\hat{\bm{k}}$ and where $k_\mrm{b}=k_0\sqrt{\epsilon_\mrm{b}}$ is
the wave number of the background medium. By expanding the plane wave in regular spherical vector waves as expressed in \eqref{eq:EHsphdef}, it can readily be shown that
\begin{equation}\label{eq:intEisquare}
\int_{V_a}\left|\bm{E}_\mrm{i}(\bm r)\right|^2\diff v  
= \left| \bm{E}_0 \right|^2 2\pi \sum_{\tau=1}^{2}\sum_{l=1}^{\infty}(2l+1)W_{\tau l}(k_\mrm{b},a),
\end{equation}
with $W_{\tau l}(k_\mrm{b},a)$ defined in \eqref{eq:Wtauldef}, and
where we have employed the orthogonality relationships \eqref{eq:vorthogonal} and \eqref{eq:Wtauldef}, as well as \eqref{eq:aiplanewave} and \eqref{eq:sumoverm}.
It is observed that for electrically small objects of size $k_0a<1$, the influence of the background medium can be appropriately neglected, and thus the incident field can be assumed to have a constant amplitude $\bm{E}_\mrm{i}(\bm r)=\bm{E}_0$. 
Hence, the relationship \eqref{eq:intEisquare} simplifies as 
\begin{equation}\label{eq:intEisquarereal}
\int_{V_a}\left|\bm{E}_\mrm{i}(\bm r)\right|^2\diff v=
\left| \bm{E}_0 \right|^2 V_\mrm{a},
\end{equation}
where $V_a=\sum_i^{N}V_i=4\pi a^3/3$ is the volume of the layered sphere.
Note that the expression in \eqref{eq:intEisquarereal} is always valid for the case with a lossless surrounding medium, where $k_\mrm{b}$ is real-valued ($k_\mrm{b}=k_\mrm{b}^*$).

The variational upper bound on the absorption cross section $\sigma_\mrm{a}^\mrm{var}$ is obtained by normalizing with 
the intensity of the plane wave at the origin $\bm{r}=\bm{0}$, \ie 
\begin{equation}\label{eq:intIi}
I_\mrm{i}=\left| \bm{E}_0 \right|^2\Re\{\sqrt{\epsilon_\mrm{b}}\}/2\eta_0,
\end{equation}
giving
\begin{equation}\label{eq:sigmaavar}
\sigma_\mrm{a}^\mrm{var}=\frac{k_0}{\Re\{\sqrt{\epsilon_\mrm{b}}\}}\frac{\alpha^2}{4}\sum_{i=1}^{N}\frac{\left|\epsilon_i-\epsilon_\mrm{b}^*\right|^2}{\Im\{\epsilon_i\}}
\int_{V_i}\left|\eu^{\mrm{i} k_\mrm{b}\hat{\bm{k}}\cdot\bm{r}} \right|^2\diff v.
\end{equation}
To give an explicit formula for \eqref{eq:sigmaavar}, it is more convenient to express the normalized absorption cross section
$Q_\mrm{a}^\mrm{var}=\sigma_\mrm{a}^\mrm{var}/\pi a^2$ for the $N$-layered sphere as
\begin{equation}\label{eq:Qavar}
Q_\mrm{a}^\mrm{var}=\frac{k_0a}{\Re\{\sqrt{\epsilon_\mrm{b}}\}}\frac{\alpha^2}{4}\sum_{i=1}^{N}\frac{\left|\epsilon_i-\epsilon_\mrm{b}^*\right|^2}{\Im\{\epsilon_i\}}
\frac{1}{\pi a^3}\int_{V_i}\left|\eu^{\mrm{i} k_\mrm{b}\hat{\bm{k}}\cdot\bm{r}} \right|^2\diff v,
\end{equation}
where 
\begin{multline}\label{eq:normabsint}
\frac{1}{\pi a^3}\int_{V_i}\left|\eu^{\mrm{i} k_\mrm{b}\hat{\bm{k}}\cdot\bm{r}} \right|^2\diff v \\
=2\sum_{\tau=1}^{2}\sum_{l=1}^{\infty}\frac{(2l+1)}{a^3}\left[W_{\tau l}(k_\mrm{b},a_i)-W_{\tau l}(k_\mrm{b},a_{i-1})\right],
\end{multline}
and where $a_i$ is the radius of each subsphere for $i=1,\ldots,N$, $a_N=a$ and $a_0=0$.
Here,
\begin{equation}\label{eq:W1l}
W_{1 l}(k_\mrm{b},a_i)=a_i^2\frac{\Im\left\{ k_\mrm{b}\mrm{j}_{l+1}( k_\mrm{b}a_i)\mrm{j}_{l}^*( k_\mrm{b}a_i)\right\}}{\Im\{k_\mrm{b}^2\}},
\end{equation}
and 
\begin{equation}\label{eq:W2l}
W_{2 l}(k_\mrm{b},a_i)=\frac{(l+1)W_{1,l-1}(k_\mrm{b},a_i)+lW_{1,l+1}(k_\mrm{b},a_i)}{(2l+1)},
\end{equation}
are readily obtained from \eqref{eq:W1ldef} and \eqref{eq:W2ldef}.
Note that for $i=1$, the last term in \eqref{eq:normabsint} vanishes because of $W_{\tau l}(k_\mrm{b},a_0)=0$ for $a_0=0$, see \eqref{eq:W1l} and \eqref{eq:W2l}, respectively.

In the case of a homogeneous ($N=1$) sphere in a lossless medium where $\Im\{\epsilon_\mrm{b}\}=0$, we have $q=0$, $\alpha=-2$ and the integral in \eqref{eq:intEisquarereal}
so that the bound in \eqref{eq:Paopt2} simplifies to
\begin{equation}\label{eq:Pudef2lossless}
P_\mrm{a}^\mrm{var}=\frac{k_0}{2\eta_0}\frac{\left|\epsilon-\epsilon_\mrm{b}\right|^2}{\Im\{\epsilon\}}\left| \bm{E}_0 \right|^2 V_a,
\end{equation}
and which reproduces the corresponding result in \cite[Eq.~(32b) on p.~3345]{Miller+etal2016}.

\section{Optical theorem based on the exterior fields}\label{sect:multbound}

\subsection{Notation and conventions}
The definition of the spherical vector waves\cite{Newton1982,Bohren+Huffman1983,Bostrom+Kristensson+Strom1991,Arfken+Weber2001,Jackson1999,Kristensson2016}
and their most important properties employed in this paper are summarized in Appendix \ref{sect:spherical}.
In particular, the regular spherical Bessel functions, the Neumann functions, the spherical Hankel functions of the first kind 
and the corresponding Riccati-Bessel functions \cite{Kristensson2016} are denoted $\mrm{j}_l(z)$, $\mrm{y}_l(z)$, $\mrm{h}_l^{(1)}(z)=\mrm{j}_l(z)+\iu\mrm{y}_l(z)$,
$\psi_l(z)=z\mrm{j}_l(z)$ and $\xi_l(z)=z\mrm{h}_l^{(1)}(z)$, respectively, all of order $l$.

\subsection{Optical theorem and physical bounds for a spherical region in a lossy medium}
We consider the physical bounds on absorption that can be derived from the optical theorem when it is
formulated in terms of the multipole coefficients of a scattering problem. In particular, the scatterer is here
embedded in a spherical region surrounded by a lossy medium, as shown in Fig.~\ref{fig:matfig11}.
Hence, the scatterer may consist of a general bianisotropic linear material and is bounded by a spherical surface of radius $a$.
The surrounding medium is an infinite homogeneous and isotropic dielectric free space having 
relative permittivity $\epsilon_\mrm{b}$ and wave number $k_\mrm{b}=k_0\sqrt{\epsilon_\mrm{b}}$. 
For simplicity, it is assumed that the background is non-magnetic (a magnetic background with relative permeability $\mu_\mrm{b}\neq 1$ can straightforwardly be added to the analysis if required). 
The background is furthermore assumed to be passive, and possibly lossy, so that $\Im\{\epsilon_\mrm{b}\}\geq 0$, and with permittivity $\epsilon_\mrm{b}$ that does not reside at the negative part of the real axis, which corresponds to the branch cut of the square root. 

The optical theorem is once again given by the power balance 
\eqref{eq:Opticaltheorem} with
the absorbed, scattered, extinct (total) and the incident powers defined by \eqref{eq:Padef} through \eqref{eq:Pidef}, respectively.
Let $a_{\tau ml}^\mrm{i}$ and $f_{\tau ml}$ denote the multipole coefficients of the incident (regular) and the scattered (outgoing) spherical vector waves, respectively,
as defined in \eqref{eq:EHsphdef}.
Based on the orthogonality of the spherical vector waves on the spherical surface $\partial V_a$ as given by \eqref{eq:orthcrosssph1} and \eqref{eq:orthcrosssph2}, it can be shown that 
\begin{eqnarray}
P_\mrm{s} & = & \frac{\Re\{\sqrt{\epsilon_\mrm{b}}\}}{2\left|k_\mrm{b}\right|^2\eta_0}\sum_{\tau,m,l}A_{\tau l}\left|f_{\tau ml}\right|^2,  \label{eq:Paexprout}\\
P_\mrm{t} & = & \frac{\Re\{\sqrt{\epsilon_\mrm{b}}\}}{2\left|k_\mrm{b}\right|^2\eta_0}\sum_{\tau,m,l}2\Re\{B_{\tau l}a_{\tau ml}^\mrm{i*}f_{\tau ml}\}, \label{eq:Ptexprout} \\
P_\mrm{i} & = & \frac{\Re\{\sqrt{\epsilon_\mrm{b}}\}}{2\left|k_\mrm{b}\right|^2\eta_0}\sum_{\tau,m,l}C_{\tau l}\left|a_{\tau ml}^\mrm{i}\right|^2, \label{eq:Piexprout}
\end{eqnarray}
where
\begin{eqnarray}
A_{\tau l} & = & \frac{1}{\Re\{k_\mrm{b}\}}\left\{
\begin{array}{ll}
-\Im\{k_\mrm{b}^*\xi_l\xi_l^{\prime *}\} & \tau=1, \vspace{0.2cm} \\
\Im\{k_\mrm{b}^*\xi_l^{\prime}\xi_l^{*}\} & \tau=2,
\end{array}
\right. \label{eq:Atauldef}
\\
B_{\tau l} & = & \frac{1}{\iu 2\Re\{k_\mrm{b}\}}\left\{
\begin{array}{ll}
k_\mrm{b}^*\xi_l\psi_l^{\prime *}-k_\mrm{b}\psi_l^*\xi_l^\prime & \tau=1, \vspace{0.2cm} \\
-k_\mrm{b}^*\xi_l^\prime\psi_l^{*}+k_\mrm{b}\psi_l^{\prime *}\xi_l & \tau=2,
\end{array}
\right.\label{eq:Btauldef}
\\
C_{\tau l} & = & \frac{1}{\Re\{k_\mrm{b}\}}\left\{
\begin{array}{ll}
\Im\{k_\mrm{b}^*\psi_l\psi_l^{\prime *}\} & \tau=1, \vspace{0.2cm} \\
-\Im\{k_\mrm{b}^*\psi_l^{\prime}\psi_l^{*}\} & \tau=2,
\end{array}
\right.\label{eq:Ctauldef}
\end{eqnarray}
for $\tau=1,2$ and $l=1,\ldots,\infty$,
and where the arguments of the Riccati-Bessel functions are $z=k_\mrm{b}a$, see also \cite[Eqs.~(8) through (11)]{Nordebo+etal2019a} and \cite[Eqs. (6) and (7)]{Sudiarta+Chylek2001}.
By applying Poynting's theorem to the scattered and the incident powers defined by \eqref{eq:Psdef} and \eqref{eq:Pidef}, it follows that $P_\mrm{s}\geq 0$ and $P_\mrm{i}\geq 0$
for a passive background medium, and hence that $A_{\tau l}> 0$ and  $C_{\tau l}\geq 0$. Note that $B_{\tau l}$ is a complex-valued constant.
For a lossless medium with $\Im\{k_\mrm{b}\}=0$, we can employ the Wronskian of the Riccati-Bessel functions $\psi_l\xi_l^\prime-\psi_l^\prime\xi_l=\iu$ and use $\xi_l^*=2\psi_l-\xi_l$
to show that the coefficients defined in \eqref{eq:Atauldef} through \eqref{eq:Ctauldef} become $A_{\tau l}=1$, $B_{\tau l}=-1/2$ and $C_{\tau l}=0$ in agreement with \eg \cite[Eq.~(7.18)]{Kristensson2016},
see also \cite[Eqs.~(9) through (12)]{Nordebo+etal2019a}.

\subsubsection{Optimal absorption of an arbitrary linear scatterer circumscribed by a sphere}
Consider the contribution to the absorbed power from a single partial wave with fixed multi-index $(\tau, m,l)$,
\begin{multline}\label{eq:Papartwave}
P_{\mrm{a},\tau ml}=\frac{\Re\{\sqrt{\epsilon_\mrm{b}}\}}{2\left|k_\mrm{b}\right|^2\eta_0}\left[-A_{\tau l}\left|f_{\tau ml}\right|^2 \right. \\
\left. +2\Re\{B_{\tau l}a_{\tau ml}^\mrm{i*}f_{\tau ml}\}+C_{\tau l}\left|a_{\tau ml}^\mrm{i}\right|^2\right],
\end{multline}
where we have employed the optical theorem \eqref{eq:Opticaltheorem} as well as \eqref{eq:Paexprout} through \eqref{eq:Piexprout}.
Let the scattering coefficients $f_{\tau ml}$ be given by the T-matrix \cite[Eq.~(7.34)]{Kristensson2016} for an arbitrary linear scatterer inside the spherical surface $\partial V_a$, so that
\begin{equation}\label{eq:T-matrixdef}
f_{n}=\sum_{n^\prime}
T_{n,n^\prime}a_{n^\prime}^\mrm{i},
\end{equation}
and where we have introduced the multi-index notation $n=(\tau, m,l)$.
It is observed that \eqref{eq:Papartwave} is a concave function of the complex-valued variables $T_{n,n^\prime}$ with respect to the primed index $n^\prime$.
Differentiating \eqref{eq:Papartwave} with respect to $T_{n,n^\prime}$ (for fixed $n$) gives the condition for stationarity
\begin{equation}\label{eq:stationarity}
A_{\tau l}a_{n^{\prime}}^\mrm{i}
\sum_{n^{\prime\prime}}a_{n^{\prime\prime}}^{\mrm{i}*}
T_{n,n^{\prime\prime}}^*=B_{\tau l}a_{n}^{\mrm{i}*}a_{n^{\prime}}^\mrm{i},
\end{equation}
which is an infinite-dimensional linear system of equations in the double-primed indices of $T_{n,n^{\prime\prime}}$. 
The corresponding system matrix $a_{n^{\prime}}^\mrm{i}a_{n^{\prime\prime}}^{\mrm{i}*}$ is of rank one and is in general unbounded.
Assuming that this matrix is either bounded, or is truncated to some finite dimension, the corresponding matrix norm is given by
\begin{equation}\label{eq:gdef}
g=\sum_{\tau,m,l}\left| a_{\tau ml}^\mrm{i}\right|^2,
\end{equation}
and the unique minimum norm (pseudo-inverse) T-matrix solution to \eqref{eq:stationarity} is given by
\begin{equation}\label{eq:Tmatsol}
T_{n,n^\prime}=
\frac{B_{\tau l}^*}{A_{\tau l}g}a_{n}^{\mrm{i}}a_{n^{\prime}}^{\mrm{i}*}.
\end{equation}
By inserting \eqref{eq:T-matrixdef} into \eqref{eq:Papartwave} and completing the squares using \eqref{eq:Tmatsol}, it can be shown that
\begin{multline}\label{eq:completesquares}
P_{\mrm{a},\tau ml}=\frac{\Re\{\sqrt{\epsilon_\mrm{b}}\}}{2\left|k_\mrm{b}\right|^2\eta_0}\Bigg\{    \\
-A_{\tau l}\left| \sum_{\tau^\prime m^\prime l^\prime}\left(T_{\tau ml,\tau^\prime m^\prime l^\prime}-
\frac{B_{\tau l}^*a_{\tau ml}^\mrm{i}a_{\tau^\prime m^\prime l^\prime}^{\mrm{i}*}}{A_{\tau l}g}\right)
a_{\tau^\prime m^\prime l^\prime}^\mrm{i}  \right|^2 \\
\left. +\left(\frac{\left| B_{\tau l} \right|^2}{A_{\tau l}}+C_{\tau l}\right)\left| a_{\tau ml}^\mrm{i}\right|^2\right\}.
\end{multline}
Due to the concavity of this expression ($A_{\tau l}>0$) it follows that the last term, which is independent of $T_{\tau ml,\tau^\prime m^\prime l^\prime}$,
gives the optimal absorption. Summing over the $\tau ml$-indices, the optimal absorption is hence obtained as
\begin{equation}\label{eq:newboundsPa1}
P_\mrm{a}^\mrm{opt}=\frac{\Re\{\sqrt{\epsilon_\mrm{b}}\}}{2\left|k_\mrm{b}\right|^2\eta_0}\sum_{\tau,m,l}
\left(\frac{\left| B_{\tau l} \right|^2}{A_{\tau l}}+C_{\tau l}\right)\left| a_{\tau ml}^\mrm{i}\right|^2.
\end{equation}

It is emphasized that the infinite dimensional matrix equation in \eqref{eq:stationarity} in general is related
to an unbounded operator where the series in \eqref{eq:gdef} does not converge (the corresponding matrix norm does not exist).
However, this is merely a mathematical subtlety that does not pose any real problem here.
Hence, considering that the T-matrix in \eqref{eq:T-matrixdef} can be truncated to a finite size $L$ with $l,l^\prime\leq L$,
the bound in \eqref{eq:newboundsPa1} can be interpreted as 
the optimal absorption with respect to all incident and scattered fields up to multipole order $L$, as $L\rightarrow\infty$.
Note in particular that the individual terms appearing in \eqref{eq:newboundsPa1} do not depend on the truncation order,
and it is only the interpretation of the partial sums that depend on $L$. 

For an incident plane wave where $\bm{E}_\mrm{i}(\bm r)=\bm{E}_0\eu^{\mrm{i} k_\mrm{b}\hat{\bm{k}}\cdot\bm{r}}$, the
multipole coefficients $a_{\tau ml}^\mrm{i}$ are given by \eqref{eq:aiplanewave}, and the optimal bound becomes
\begin{equation}\label{eq:newboundsPa2}
P_\mrm{a}^\mrm{opt}=\frac{\pi\Re\{\sqrt{\epsilon_\mrm{b}}\}\left| \bm{E}_0\right|^2}{\left|k_\mrm{b}\right|^2\eta_0}\sum_{\tau=1}^{2}\sum_{l=1}^\infty(2l+1)
\left(\frac{\left| B_{\tau l} \right|^2}{A_{\tau l}}+C_{\tau l}\right),
\end{equation}
where we have made use of the sum identities~\eqref{eq:sumoverm} for the vector spherical harmonics.
The corresponding optimal normalized absorption cross section $Q_\mrm{a}^\mrm{opt}$ is obtained by normalizing with 
the intensity $I_\mrm{i}$ of the plane wave at the origin $\bm{r}=\bm{0}$ given by \eqref{eq:intIi}, as well as with
the geometrical area cross section of the sphere $\pi a^2$, giving
\begin{equation}\label{eq:newboundQa}
Q_\mrm{a}^\mrm{opt}=\frac{2}{\left|k_\mrm{b}a\right|^2}\sum_{\tau=1}^{2}\sum_{l=1}^\infty(2l+1)
\left(\frac{\left| B_{\tau l} \right|^2}{A_{\tau l}}+C_{\tau l}\right).
\end{equation}

\begin{figure}[htb]
\begin{center}
\includegraphics[width=0.48\textwidth]{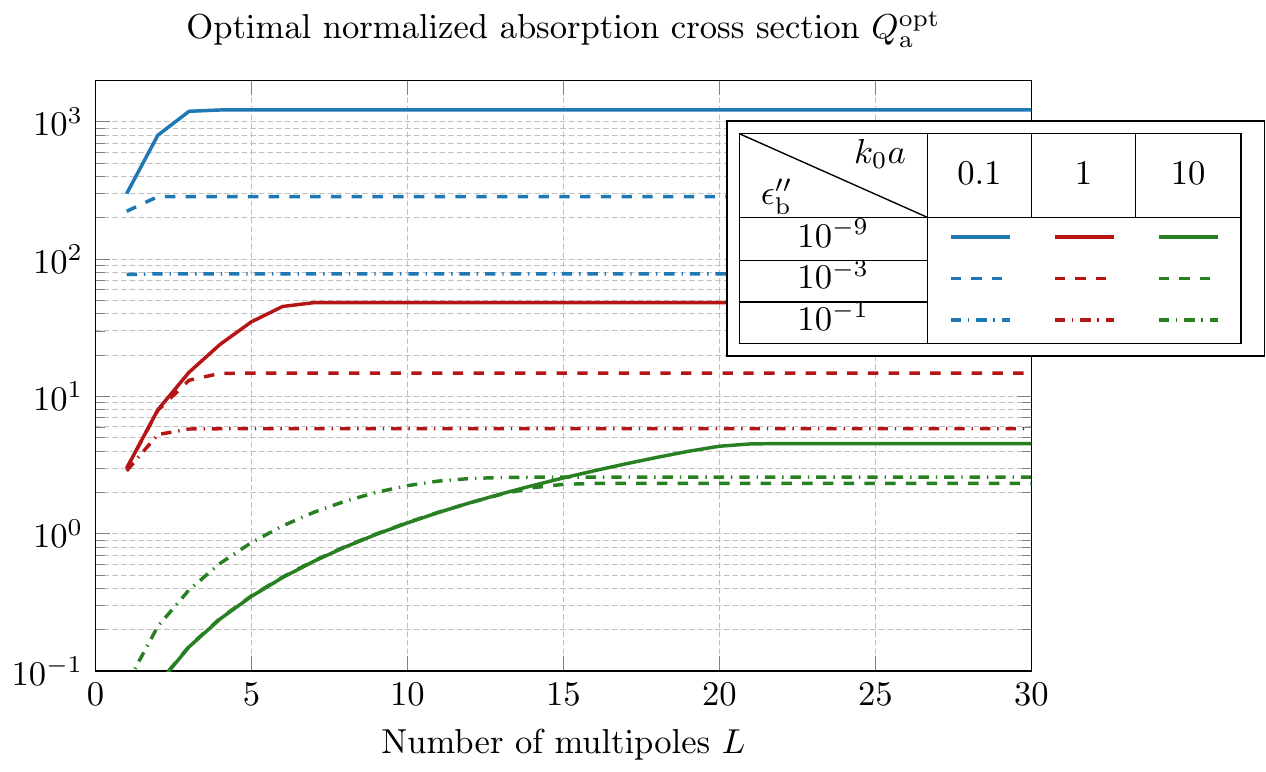}
\end{center}
\vspace{-5mm}
\caption{Optimal normalized absorption cross section $Q_\mrm{a}^\mrm{opt}$ of a sphere in a lossy medium,
plotted as a function of the number of included multipoles $L$. }
\label{fig:matfig801}
\end{figure}

In the next section we will show that \eqref{eq:newboundQa} converges whenever there are losses in the exterior medium and $\Im\{k_\mrm{b}\}> 0$.
In Fig.~\ref{fig:matfig801} is illustrated the convergence of the expression \eqref{eq:newboundQa} by plotting the partial sums against the number of included multipoles $L$.
The calculations are for electrical sizes $k_0a\in\{0.1,1,10\}$ and with background losses $\epsilon_\mrm{b}^{\prime\prime}\in\{10^{-9},10^{-3},10^{-1}\}$
where $\epsilon_\mrm{b}=1+\iu\epsilon_\mrm{b}^{\prime\prime}$. Clearly, with increasing external losses there are fewer modes that can contribute to the absorption inside the sphere,
which is due to the interaction of the reactive near-fields of the higher order modes with the lossy exterior domain.

In the lossless case, when $\Im\{k_\mrm{b}\}=0$, the truncated partial sums of \eqref{eq:newboundQa} can be calculated as
\begin{equation}\label{eq:newboundQaLL}
Q_{\mrm{a}}^{\mrm{opt},L}=\frac{2}{(k_\mrm{b}a)^2}\sum_{\tau=1}^{2}\sum_{l=1}^L
\frac{2l+1}{4}=\frac{1}{(k_\mrm{b}a)^2}L(L+2),
\end{equation}
where $L$ is the truncated maximal multipole order, which can be determined by the method proposed in \cite[p.~1508]{Wiscombe1980}. Interestingly, the obtained result in \eqref{eq:newboundQaLL} is similar as the expression for maximum gain derived by Harrington in \cite[Eq.~(11) on p.~221]{Harrington1958}. 

\subsubsection{Proof of convergence}
To prove that \eqref{eq:newboundQa} converges for a lossy medium where $\Im\{k_\mrm{b}\}> 0$, we consider the following power series expansions of the
regular Riccati-Bessel functions 
\begin{equation}\label{eq:psiseries}
\psi_l(z)=\sum_{k=0}^\infty \alpha_{kl}z^{l+1+2k},
\end{equation}
and the singular Riccati-Hankel functions
\begin{equation}\label{eq:xiseries}
\xi_l(z)=\iu\sum_{k=0}^{l} \beta_{kl}z^{-l+2k}+\alpha_{0l}z^{l+1}+{\cal O}\{z^{l+2}\},
\end{equation}
where $\alpha_{kl}=(-1/2)^k/k!(2l+2k+1)!!$ and $\beta_{kl}=-(1/2)^k(2l-2k-1)!!/k!$ \cf \cite[Eqs.~(10.53.1) and (10.53.2)]{Olver+etal2010}
and where ${\cal O}\{\cdot\}$ denotes the big ordo defined in \cite[p.~4]{Olver1997}. 
By inserting \eqref{eq:psiseries} and \eqref{eq:xiseries} into \eqref{eq:Atauldef} through  \eqref{eq:Ctauldef} and retaining only the most dominating terms for fixed $z=k_\mrm{b}a$ and increasing $l$, it is found that
\begin{equation}\label{eq:Ataulasympt}
A_{\tau l} \sim \left\{
\begin{array}{ll}
\displaystyle \frac{2\beta_{0l}\beta_{1l}\sin 2\theta}{|z|^{2l-1}\cos\theta} & \tau=1, \vspace{0.2cm} \\
\displaystyle \frac{l\beta_{0l}^2\sin 2\theta}{|z|^{2l+1}\cos\theta} & \tau=2,
\end{array}
\right.
\end{equation}
\begin{equation}\label{eq:Btaulasympt}
B_{\tau l} \sim \left\{
\begin{array}{ll}
\displaystyle -\frac{1}{2}\frac{\eu^{-\iu\theta(2l+1)}}{\cos\theta} & \tau=1, \vspace{0.2cm} \\
\displaystyle -\frac{1}{2}\frac{l\eu^{-\iu\theta(2l+3)}+(l+1)\eu^{-\iu\theta(2l-1)}}{(2l+1)\cos\theta}  & \tau=2,
\end{array}
\right.
\end{equation}
\begin{equation}\label{eq:Ctaulasympt}
C_{\tau l} \sim \left\{
\begin{array}{ll}
\displaystyle \frac{-2\alpha_{0l}\alpha_{1l}|z|^{2l+3}\sin 2\theta}{\cos\theta} & \tau=1, \vspace{0.2cm} \\
\displaystyle \frac{(l+1)\alpha_{0l}^2|z|^{2l+1}\sin 2\theta}{\cos\theta} & \tau=2,
\end{array}
\right.
\end{equation}
and where $z=|z|\eu^{\iu \theta}$. Note that
$\alpha_{0l}=1/(2l+1)!!$, $\alpha_{1l}=-(1/2)/(2l+3)!!$, $\beta_{0l}=-(2l-1)!!$ and $\beta_{1l}=-(1/2)(2l-3)!!$.
Convergence of \eqref{eq:newboundQa} can now be established by considering the
factorial increase of the $A_{\tau l}$ coefficients, the boundedness of the $B_{\tau l}$ coefficients
and the factorial decrease of the $C_{\tau l}$ coefficients for large $l$. In the lossless case when
$\theta=0$, we have $A_{\tau l}=1$, $B_{\tau l}=-1/2$ and $C_{\tau l}=0$ and \eqref{eq:newboundQa} 
is divergent as demonstrated in \eqref{eq:newboundQaLL}.

\section{Numerical examples}\label{sect:numexamples}
In this section, we illustrate the theory that has been developed in Sections~\ref{sect:varbound} and \ref{sect:multbound} in comparison with the normalized absorption cross sections of spherical objects embedded in a lossy medium. 
As objects of study, homogeneous and layered (core-shell) spheres are selected. 
The dielectric background medium is characterized by permittivity $\epsilon_\mrm{b}=\epsilon_\mrm{b}^\prime+\iu\epsilon_\mrm{b}^{\prime\prime}$, 
where the choice of $\epsilon_\mrm{b}^{\prime\prime}$ is based on the skin depth of human skin $\alpha=2k_0\epsilon_\mrm{b}^{\prime\prime}$, $\alpha^{-1}\in(10^{-4},10^{-2})$~cm, see \cite[Table~3.2 on p.~49]{Duck1990}. Note that the real part of the background permittivity does not play the key role in comparisons presented below, and thus we consistently choose $\epsilon_\mrm{b}^\prime=1$ despite that the refractive index of human tissue $n\approx 1.33$ \cite[Table~3.8 on p.~63]{Duck1990}. In addition to this investigation, the absorption of spherical objects embedded in an almost lossless medium with relative permittivity $\epsilon_\mrm{b}=1+\iu 10^{-9}$ is also considered.

\begin{figure}[htb!]
\begin{center}
\includegraphics[width=0.48\textwidth]{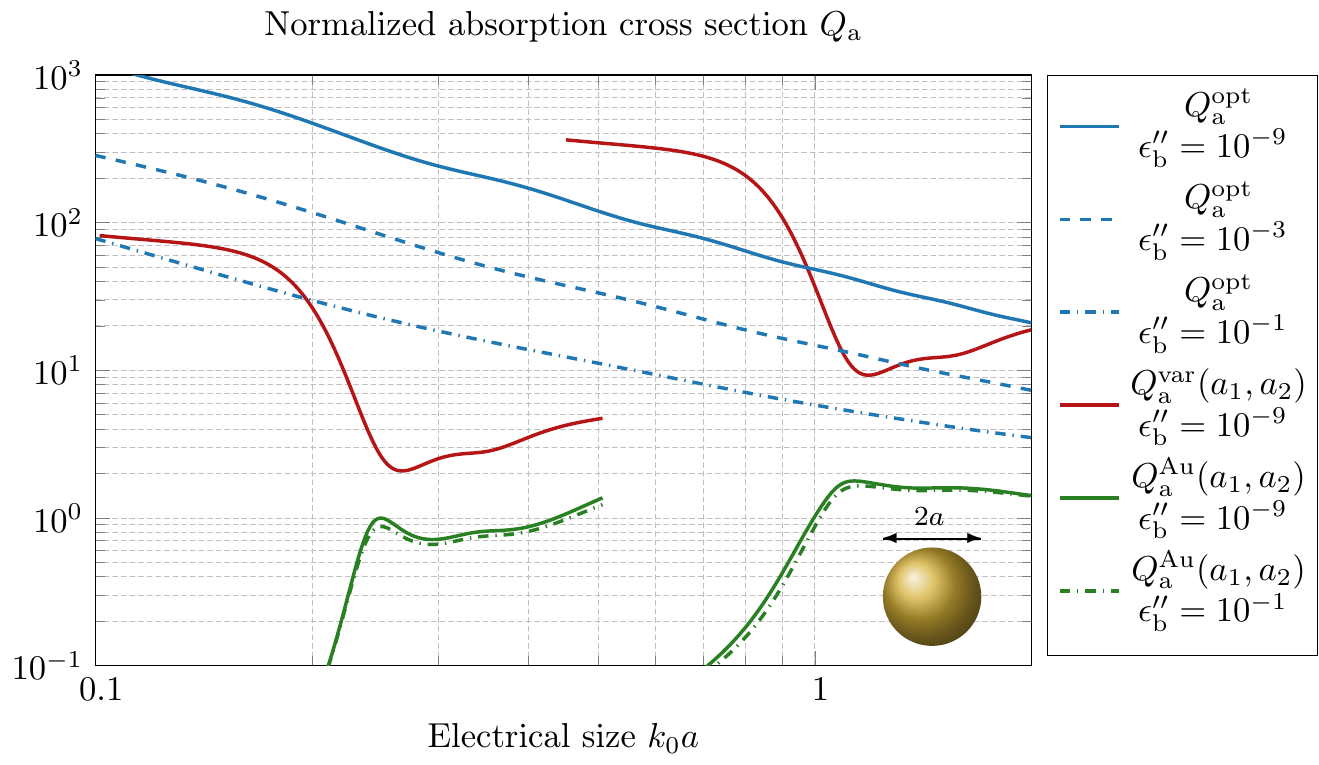}
\end{center}
\vspace{-5mm}
\caption{Comparison between the optimal normalized absorption cross section $Q_\mrm{a}^\mrm{opt}$, the 
absorption of a homogeneous sphere made of gold $Q_\mrm{a}^\mrm{Au}$, and the corresponding variational bound $Q_\mrm{a}^\mrm{var}$;
all plotted as functions of the electrical size $k_0a$.
The plots are for various levels of background loss $\epsilon_\mrm{b}^{\prime\prime}$, and the calculations of $Q_\mrm{a}^\mrm{var}$ and $Q_\mrm{a}^\mrm{Au}$ are for 
two different radii of the sphere $a_1=20\unit{nm}$ (to the left) and $a_2=89\unit{nm}$ (to the right). }
\label{fig:matfig802}
\end{figure}

In Fig.~\ref{fig:matfig802} is shown a comparison of the optimal normalized absorption cross section $Q_\mrm{a}^\mrm{opt}$ given by \eqref{eq:newboundQa},
the absorption of a homogeneous sphere made of gold $Q_\mrm{a}^\mrm{Au}$ (full Mie solution) obtained by normalization of \eqref{eq:Papartwave} with the intensity $I_\mrm{i}$ of the plane wave at the origin \eqref{eq:intIi} and the geometrical area cross section $\pi a^2$, and the corresponding variational bound $Q_\mrm{a}^\mrm{var}$ for a homogeneous object given by \eqref{eq:Qavar}.
The background relative permittivity is  $\epsilon_\mrm{b}=1+\iu\epsilon_\mrm{b}^{\prime\prime}$ with various levels of background loss: $\epsilon_\mrm{b}^{\prime\prime}\in\{10^{-9},10^{-3},10^{-1}\}$.
The calculations of $Q_\mrm{a}^\mrm{Au}$ and $Q_\mrm{a}^\mrm{var}$ are for two different radii of gold spheres ($20\unit{nm}$ and $89\unit{nm}$)
and the same photon energy range 1--5\unit{eV} (corresponds to the wavelength range 248--1240~\unit{nm}) according to the Brendel-Bormann (BB) model fitted to experimental data as in 
\cite[the dielectric model in Eq. (11) with parameter values from Table 1 and Table 3]{Rakic+etal1998}.
The sphere of radius $a=89\unit{nm}$ has been tuned to optimal electric-dipole absorption for a lossless background as in \cite[Fig.~4]{Nordebo+etal2019a}.
It is noted that the variational bound  $Q_\mrm{a}^\mrm{var}$ depends very weakly on the background loss for these parameter ranges, and the
bound is therefore plotted only for $\epsilon_\mrm{b}^{\prime\prime}=10^{-9}$ (the plots for  $\epsilon_\mrm{b}^{\prime\prime}\in\{10^{-3},10^{-1}\}$ almost coincide).
As can be seen in this plot, the two bounds $Q_\mrm{a}^\mrm{opt}$ and $Q_\mrm{a}^\mrm{var}$, which are derived under different assumptions 
(arbitrary structure of linear bianisotropic materials inside the sphere vs arbitrary structure of gold inside the sphere), give complementary information
about the upper bounds on absorption. At the same time, the normalized absorption cross section $Q_\mrm{a}^\mrm{Au}$ is not tight 
with respect to the optimal bounds as shown in Fig.~\ref{fig:matfig802}, despite it is known that its electric-dipole contribution approaches the bound for electric-dipole absorption when the radius of object is 89~nm, see \cite[Fig.~4]{Nordebo+etal2019a}.

Now, we would like to find such spherical objects which are resonant at small electrical size, but at the same time are of a reasonable physical size and have a resonance absorption peak close to the optimal absorption bound. To fit these requirements, one way is to ``tune" a homogeneous sphere to the resonance at the desirable electrical size. An alternative approach is to consider a layered sphere constructed of a dielectric core and coated with a metallic shell.

\begin{figure}[htb!]
\begin{center}
\includegraphics[width=0.48\textwidth]{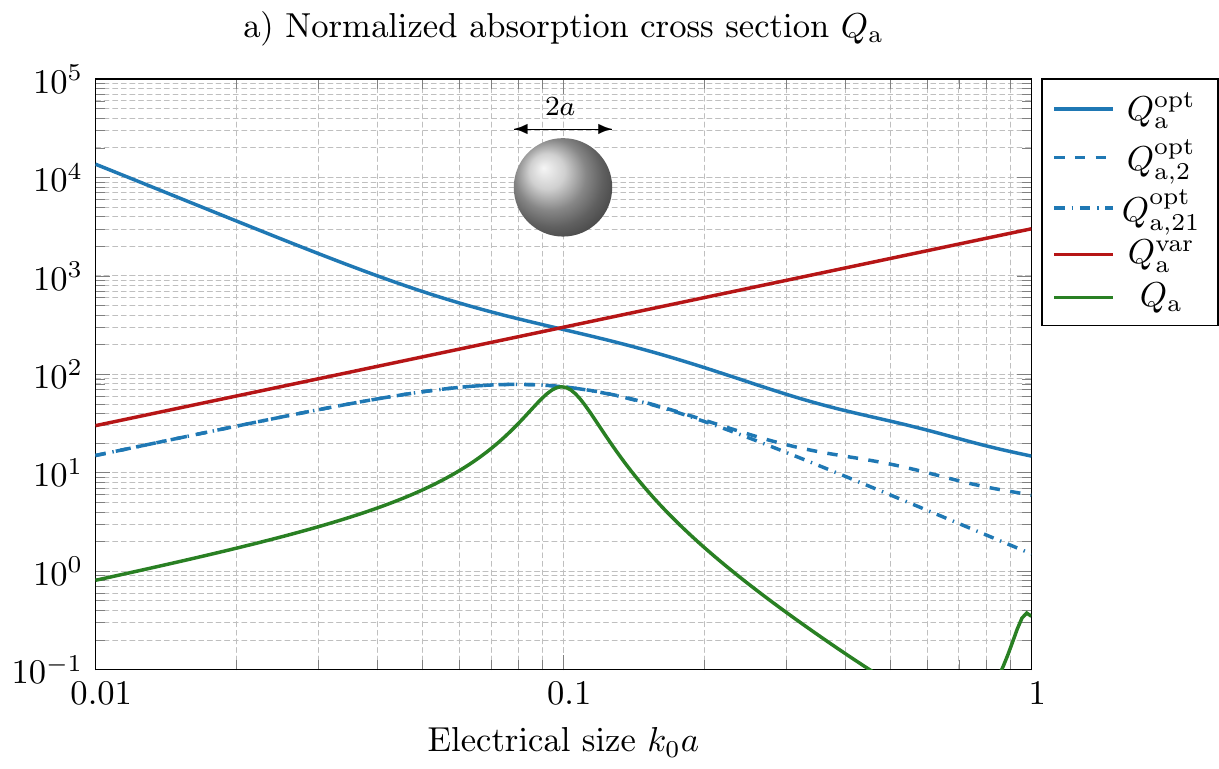}\\
\includegraphics[width=0.48\textwidth]{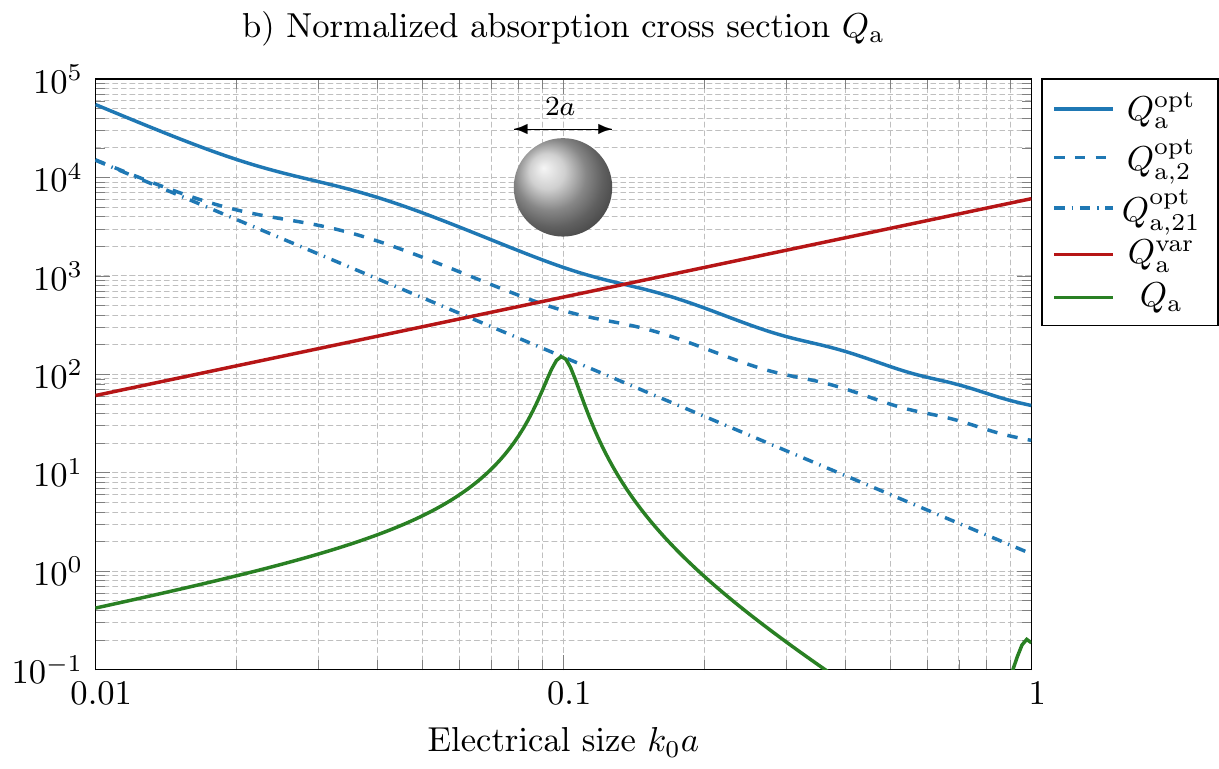}
\end{center}
\vspace{-5mm}
\caption{Comparison of the various upper bounds $Q_\mrm{a}^\mrm{opt}$,  $Q_{\mrm{a},2}^\mrm{opt}$ (the electric multipole contribution), $Q_{\mrm{a},21}^\mrm{opt}$ (the electric-dipole upper bound),
$Q_\mrm{a}^\mrm{var}$ and the absorption of a sphere $Q_\mrm{a}$ tuned to optimal electric (plasmonic) dipole resonance at $k_0a=0.1$ for: a) $\epsilon_\mrm{b}^{\prime\prime}=10^{-3}$ where $\epsilon_\mrm{b}=1+\iu\epsilon_\mrm{b}^{\prime\prime}$, and
the permittivity of the sphere is $\epsilon=-2.024+\iu 0.0040$; b) $\epsilon_\mrm{b}^{\prime\prime}=10^{-9}$ 
where $\epsilon_\mrm{b}=1+\iu\epsilon_\mrm{b}^{\prime\prime}$, and the permittivity of the sphere is $\epsilon=-2.024 + \iu 0.0020$.}
\label{fig:matfig803}
\end{figure}

In Fig.~\ref{fig:matfig803}a is shown a comparison of the various upper bounds on absorption and the absorption of a sphere tuned to optimal electric (plasmonic) dipole resonance.
Here, $Q_\mrm{a}$ denotes the full Mie solution for a homogeneous sphere with a (hypothetical) fixed value of permittivity
$\epsilon=-2\epsilon_\mrm{b}^*-\frac{12}{5}\epsilon_\mrm{b}^{*2}(0.1)^2 +\iu 2\epsilon_\mrm{b}^{*2}\sqrt{\epsilon_\mrm{b}^*}(0.1)^3$
 which has been tuned to optimal electric-dipole resonance at $k_0a=0.1$, \cf \cite[Eq.~(55)]{Nordebo+etal2019a}.
The optimal normalized absorption cross section $Q_\mrm{a}^\mrm{opt}$ is given by  \eqref{eq:newboundQa}, 
$Q_{\mrm{a},2}^\mrm{opt}$ and $Q_{\mrm{a},21}^\mrm{opt}$ denote the corresponding electric multipole contribution ($\tau=2,~l=1,2,3,\ldots$) and the normalized electric-dipole absorption cross section ($\tau=2,~l=1$), respectively. 
The corresponding variational bound $Q_\mrm{a}^\mrm{var}$ is given by  \eqref{eq:Qavar}.
All calculations have been made for $\epsilon_\mrm{b}^{\prime\prime}=10^{-3}$ where $\epsilon_\mrm{b}=1+\iu\epsilon_\mrm{b}^{\prime\prime}$.
As can be seen in Fig.~\ref{fig:matfig801}, for this combination of electrical size and background loss, it is (almost) sufficient to consider the
dipole ($l=1$) contribution at resonance, which explains why the Mie solution $Q_\mrm{a}$ in Fig.~\ref{fig:matfig803}a is (almost) tight with the upper bounds $Q_{\mrm{a},21}^\mrm{opt}$ and $Q_{\mrm{a},2}^\mrm{opt}$, respectively.
In Fig.~\ref{fig:matfig803}b is shown the same calculations, except that here the background loss is given by $\epsilon_\mrm{b}^{\prime\prime}=10^{-9}$.
Again, as can be seen in Fig.~\ref{fig:matfig801}, with such small background losses the optimal $Q_\mrm{a}^\mrm{opt}$ is based on at least multipole orders up to $L=3$,
which explains why the Mie solution $Q_\mrm{a}$ in Fig.~\ref{fig:matfig803}b is not tight with the corresponding upper bound $Q_{\mrm{a},2}^\mrm{opt}$. Interestingly that at the same time, the Mie solution can approach the upper electric dipole bound $Q_{\mrm{a},21}^\mrm{opt}$ when such an amount of losses in the background takes place.

\begin{figure}[htb]
\begin{center}
\includegraphics[width=0.48\textwidth]{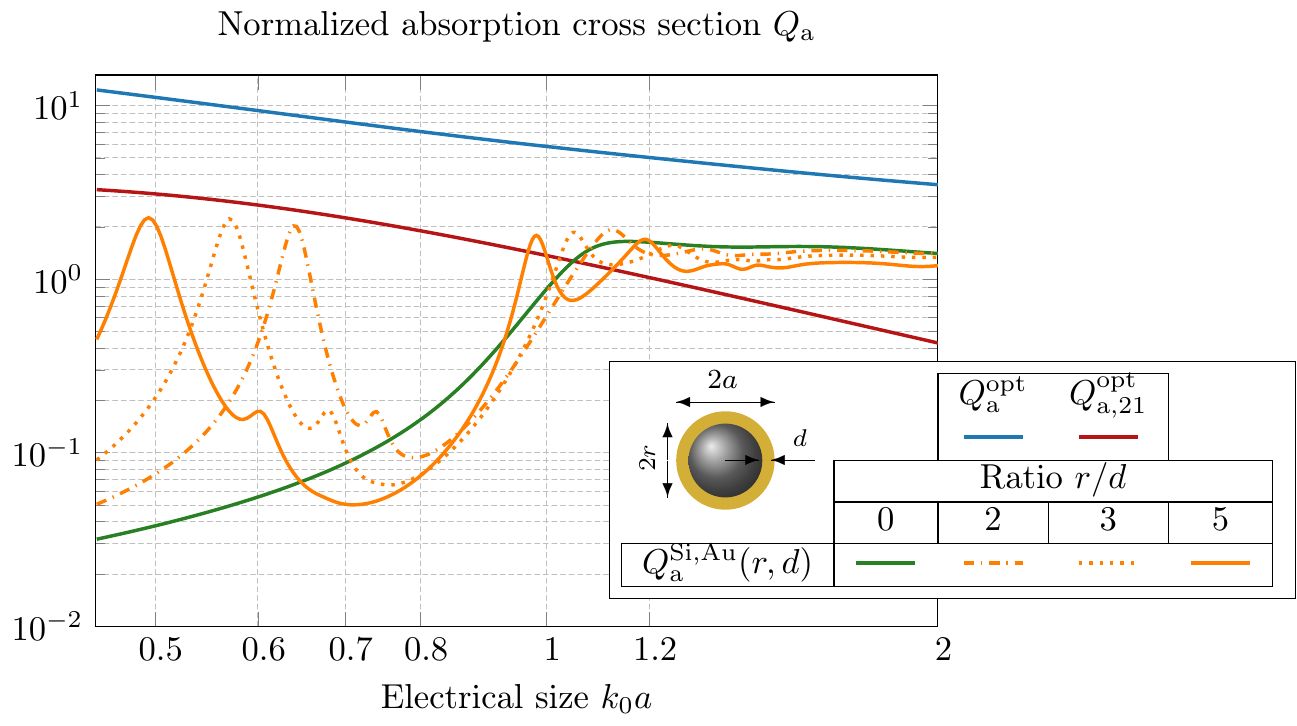}
\end{center}
\vspace{-5mm}
\caption{Comparison between the optimal total normalized absorption cross section $Q_\mrm{a}^\mrm{opt}$, the total absorption of a sphere made of gold $Q_\mrm{a}^\mrm{Au}$ (ratio $r/d=0$), and the total absorption of a multilayered sphere made of silicon (core of radius $r$) and gold (shell of thickness $d$) $Q_\mrm{a}^\mrm{Si,Au}$; 
all are plotted as functions of the electrical size $k_0a$. The plots are made for a fixed level of a background loss $\epsilon_\mrm{b}^{\prime\prime}=10^{-1}$ and a fixed total radius of spheres $a=89$ nm. }
\label{fig:matfig551}
\end{figure}

Fig.~\ref{fig:matfig551} depicts a comparison of the normalized absorption cross section $Q_\mrm{a}^\mrm{opt}$ and its electric-dipole component $Q_\mrm{a,21}^\mrm{opt}$ with the total absorption $Q_\mrm{a}$ of spherical objects of the total radius $a=89$~nm. In this plot, the following objects have been considered: a homogeneous sphere made of gold $Q_\mrm{a}^\mrm{Au}$ (special case with ratio $r/d=0$) and three designs of a layered sphere, where the core is made of silicon, and it is coated by gold. The absorption of the layered sphere is based on the normalization of \eqref{eq:Papartwave} similarly as in the previous examples, but here, the scattering coefficients $f_{\tau ml}$ \eqref{eq:T-matrixdef} are expressed in terms of the transition matrices $t_{\tau l}^{(i)}$ for layered spherical objects, see \eqref{eq:ttaulApp2} in Appendix~\ref{sect:Mietheory}. The designs have been considered for three different ratios between the radius of core $r$ and the thickness of shell $d$: $r/d\in\{2,3,5\}$. Note that $r$ and the total radius $a$ coincide with $a_1$ and $a_2$ ($N=2$), respectively, introduced in Appendix~\ref{sect:Mietheory}. The dielectric properties of silicon are represented by Drude-Lorentz model that fits the measurement data and valid in the photon range $1-6$~eV, see \cite[the dielectric model in Eq.~(4) with parameter values in Table~1]{Sehmi+etal2017}. The permittivity of background is $\epsilon_\mrm{b}=1+\iu10^{-1}$. As can be seen from Fig.~\ref{fig:matfig551}, by replacing a part of the metallic sphere with silicon, it is possible to obtain a plasmonic resonance at smaller electrical sizes $k_0a$. This is a magnetic dipole resonance, which is inherent in dielectric materials \cite{Krasnok+etal2012}. It should be noted that by increasing the ratio between the  radius of the silicon core and the thickness of the gold shell, the composed sphere becomes resonant at smaller electrical sizes: \eg for $r/d=5$, the layered sphere is resonant at $k_0a\approx 0.49$, while the sphere with $r/d=2$ and the gold sphere are resonant at $k_0a\approx 0.64$ and $k_0a\approx 1.13$, respectively. Hence, by increasing the ratio between the dielectric and metallic sizes, the resonance of the composed object will move towards the resonance of the dielectric sphere that occurs when the wavelength inside the sphere approximately equals to its diameter~\cite{Krasnok+etal2012,Miroshnichenko+etal2011}.

\begin{figure}[htb!]
\begin{center}
\includegraphics[width=0.35\textwidth]{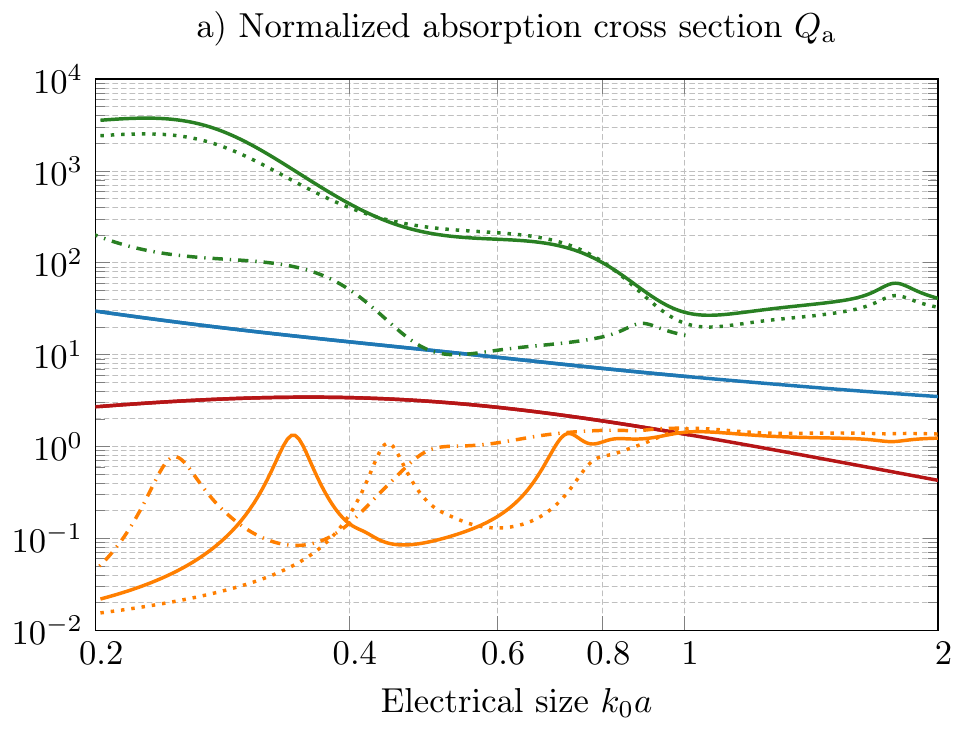} \\
\includegraphics[width=0.35\textwidth]{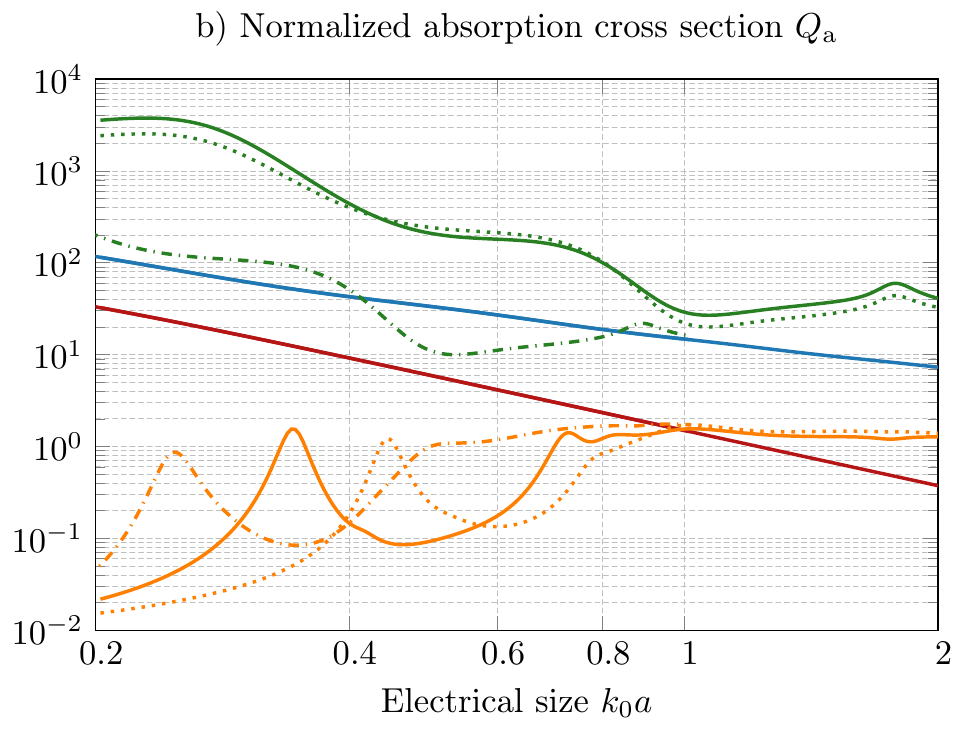} \\
\includegraphics[width=0.35\textwidth]{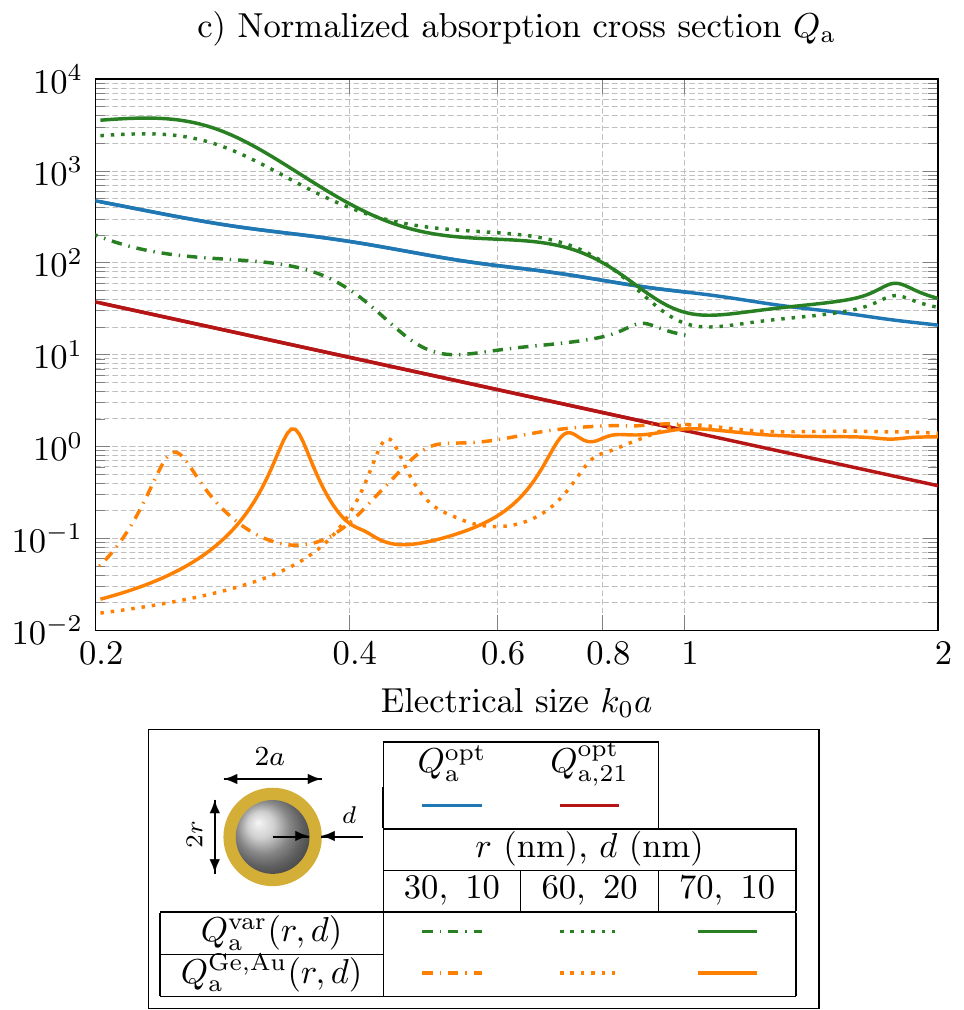}
\end{center}
\vspace{-5mm}
\caption{Comparison between the optimal total normalized absorption cross section $Q_\mrm{a}^\mrm{opt}$, variational bound on the normalized absorption cross section $Q_\mrm{a}^\mrm{var}$, and total absorption of a multilayered sphere made of germanium (core of radius $r$) 
and gold (shell of thickness $d$) 
$Q_\mrm{a}^\mrm{Ge,Au}$. All the results are plotted as functions of the electrical size $k_0a$ for different levels of losses in the background medium with relative permittivity $\epsilon_\mrm{b}=1+\iu\epsilon_\mrm{b}^{\prime\prime}$: a) $\epsilon_\mrm{b}^{\prime\prime}=10^{-1}$; b) $\epsilon_\mrm{b}^{\prime\prime}=10^{-3}$; c) $\epsilon_\mrm{b}^{\prime\prime}=10^{-9}$.}
\label{fig:matfig601-621}
\end{figure}

In Figs.~\ref{fig:matfig601-621}a-c is shown a comparison of the various upper bounds on absorption, and the absorption of a 2-layered core-shell sphere of different designs for three different levels of losses in the background: $\epsilon_\mrm{b}^{\prime\prime}\in\{10^{-1},10^{-3},10^{-9}\}$, where $\epsilon_\mrm{b} = 1+\iu\epsilon_\mrm{b}^{\prime\prime}$. The considered designs are with cores of radii $r\in\{30,60,70\}$~nm and the corresponding thicknesses of shells $d\in\{10,20,10\}$~nm. The core is made of germanium (Ge), which is characterized by Drude-Lorentz model \cite[the dielectric model in Eq.~(4) with parameter values in Table~1]{Sehmi+etal2017} valid in the photon energy range $0.5-6.0$~eV. The dielectric core is coated by gold (Au) which is characterized by the BB model \cite{Rakic+etal1998}. The corresponding variational upper bound for the 2-layered sphere $Q_\mrm{a}^\mrm{var}$ has been obtained by \eqref{eq:Qavar}. As can be seen from these figures, the composed structure based on Ge and Au is resonant at small electrical sizes $k_0a$, but the magnitude of these resonances is not tight to none of the multipole upper bounds, even when the amount of the background losses is high, see Fig.~\ref{fig:matfig601-621}a with results for $\epsilon_\mrm{b}^{\prime\prime}=10^{-1}$.  
By comparison of the results on the upper bounds for absorption $Q_\mrm{a}^\mrm{opt}$ obtained by \eqref{eq:newboundQa} and $Q_\mrm{a}^\mrm{var}$ for different amounts of losses in the background in Figs.~\ref{fig:matfig601-621}a-c, it should be noted that these bounds provide a complementary information on absorption, and thus this conclusion is valid both for homogeneous and layered spherical objects. It can be concluded that for backgrounds with strong losses, the multipole bound $Q_\mrm{a}^\mrm{opt}$ brings more information on absorption limitations, while the variational bound $Q_\mrm{a}^\mrm{var}$ is more tight for smaller objects that are embedded in low loss surrounding media, which complements the results obtained by $Q_\mrm{a}^\mrm{opt}$, see \eg $Q_\mrm{a}^\mrm{var}(r,d)$ for $r=30\unit{nm}$ and $d=10\unit{nm}$ in Fig.~\ref{fig:matfig601-621}c.

\section{Summary and conclusions}\label{sect:summaryandconclusions}
In this paper, two fundamental multipole bounds on absorption of scattering objects embedded in a lossy surrounding medium have been derived. The derivation of these bounds have been made under two fundamentally different assumptions: based on equivalent currents inside the scatterer, and with respect to the external fields using the T-matrix parameters, respectively. The first bound depends on the material properties of scatterer as well as on its shape, while the second bound is applicable to spherical objects made of an arbitrary material. Through the numerical examples, it has been illustrated that the derived bounds can complement each other, depending on the amount of losses in the surrounding medium.

\begin{acknowledgments}
This work has been supported by the Swedish Foundation for Strategic Research (SSF), 
grant no. AM13-0011 under the program Applied Mathematics and the project Complex Analysis and Convex Optimization for EM Design.
\end{acknowledgments}

\appendix

\section{Derivations based on calculus of variation}\label{sect:varcalculus}
\subsection{Optimal power absorption}\label{sect:maxPabs}
Consider the Lagrangian functional for the optimization problem \eqref{eq:Paoptdef} which is 
given by
\begin{multline}\label{eq:Lagrangian}
{\cal L}(\bm{F},\lambda)=(1-\lambda)\Im\left\{\int_{V}\bm{F}^*\cdot\bm{M}_\mrm{a}\cdot\bm{F}\diff v\right\} \\
+\lambda\Im\left\{\int_{V}\bm{F}_\mrm{i}^*\cdot\bm{M}_\mrm{t}\cdot\bm{F}\diff v\right\} \\
-\lambda\Im\left\{\int_{V}\bm{F}_\mrm{i}^*\cdot\bm{M}_\mrm{b}\cdot\bm{F}_\mrm{i}\diff v\right\},
\end{multline}
where $\lambda$ is the Lagrange multiplier. Taking the first variation of \eqref{eq:Lagrangian} yields
\begin{multline}\label{eq:diffLagrangian}
\delta\!{\cal L}(\bm{F},\lambda)=\Im\left\{\int_{V}\delta\!\bm{F}^*\cdot
\Big[(1-\lambda)\left(\bm{M}_\mrm{a}-\bm{M}_\mrm{a}^\dagger\right)\cdot\bm{F} \right. \\
\left. -\lambda\bm{M}_\mrm{t}^\dagger\cdot\bm{F}_\mrm{i} \right]\diff v
\bigg\},
\end{multline}
where 
$(\cdot)^\dagger$ denotes the Hermitian transpose. 
Hence, a stationary solution with $\delta\!{\cal L}(\bm{F},\lambda)=0$ is given by
\begin{equation}\label{eq:stationarysol}
\bm{F}=\frac{\alpha}{2\iu}\left(\Im\{\bm{M}_\mrm{a}\} \right)^{-1}\cdot\bm{M}_\mrm{t}^\dagger\cdot\bm{F}_\mrm{i},
\end{equation}
where $\alpha=\lambda/(1-\lambda)$, and 
\begin{equation}
\Im\{\bm{M}_\mrm{a}\}=\frac{\bm{M}_\mrm{a}-\bm{M}_\mrm{a}^\dagger}{2\iu}.
\end{equation}
Inserting the solution \eqref{eq:stationarysol} into \eqref{eq:Paexpr} gives the optimal absorption
\begin{equation}\label{eq:PaoptApp_res}
P_\mrm{a}^\mrm{opt}=\frac{k_0\alpha^2}{8\eta_0}\int_{V}
\bm{F}_\mrm{i}^*\cdot\bm{M}_\mrm{t}\cdot\left(\Im\{\bm{M}_\mrm{a}\} \right)^{-1}\cdot\bm{M}_\mrm{t}^\dagger\cdot\bm{F}_\mrm{i}\diff v.
\end{equation}

\subsection{Maximization of parameter $q$}\label{sect:minq}
Consider the function $f(\bm{M}_\mrm{a})$
\begin{equation}\label{eq:fMadef}
f(\bm{M}_\mrm{a})=\int_V \bm{F}_\mrm{i}^*\cdot\bm{M}_\mrm{t}\cdot\left( \Im\{\bm{M}_\mrm{a}\} \right)^{-1}\cdot\bm{M}_\mrm{t}^\dagger\cdot\bm{F}_\mrm{i}\diff v,
\end{equation}
which represents the denominator of parameter $q$ defined in \eqref{eq:qdef}, and where $\bm{M}_\mrm{t} = \bm{M}_\mrm{a} - \bm{M}_\mrm{b}^\dagger$ is based on definitions \eqref{eq:Madef} through \eqref{eq:Mtdef}. The function \eqref{eq:fMadef} is convex in $\bm{M}_\mrm{a}$ for $\Im\{\bm{M}_\mrm{a}\}>0$, and hence in order to maximize $q$ in \eqref{eq:qdef}, $f(\bm{M}_\mrm{a})$ has to be minimized. The first variation of \eqref{eq:fMadef} is given by
\begin{multline}\label{eq:delfMadef}
\hspace{-0.25cm}\delta f(\bm{M}_\mrm{a}) = \int_{V}\bm{F}_\mrm{i}^*\cdot\delta\bm{M}_\mrm{a}\cdot\left( \Im\{\bm{M}_\mrm{a}\} \right)^{-1}\cdot\left( \bm{M}_\mrm{a}-\bm{M}_\mrm{b}^\dagger \right)^\dagger\cdot\bm{F}_\mrm{i}\diff v \\
- \int_{V}\bm{F}_\mrm{i}^*\cdot\left( \bm{M}_\mrm{a}-\bm{M}_\mrm{b}^\dagger \right)
\cdot\left( \Im\{\bm{M}_\mrm{a}\} \right)^{-1}\cdot\Im\{\delta\bm{M}_\mrm{a}\} \\
\cdot\left( \Im\{\bm{M}_\mrm{a}\} \right)^{-1}\cdot\left( \bm{M}_\mrm{a}-\bm{M}_\mrm{b}^\dagger \right)^\dagger\cdot\bm{F}_\mrm{i}\diff v \\
+\int_{V}\bm{F}_\mrm{i}^*\cdot\left( \bm{M}_\mrm{a}-\bm{M}_\mrm{b}^\dagger \right)
\cdot\left( \Im\{\bm{M}_\mrm{a}\} \right)^{-1}\cdot\delta\bm{M}_\mrm{a}^\dagger\cdot\bm{F}_\mrm{i}\diff v,
\end{multline}
and where we have employed the relation 
\begin{multline}\label{eq:deltaMainv}
\delta\left( \Im\{\bm{M}_\mrm{a}\} \right)^{-1} \\
=-\left( \Im\{\bm{M}_\mrm{a}\} \right)^{-1}\cdot\Im\{\delta\bm{M}_\mrm{a}\}\cdot\left( \Im\{\bm{M}_\mrm{a}\} \right)^{-1}.
\end{multline}
The stationarity condition $\delta f(\bm{M}_\mrm{a})=0$ for all $\delta\bm{M}_\mrm{a}$ and all $\bm{F}_\mrm{i}$ gives the simplified condition
\begin{multline}
\Re\left\{ \delta\bm{M}_\mrm{a}\cdot\left( \Im\{\bm{M}_\mrm{a}\} \right)^{-1}\cdot\left( \bm{M}_\mrm{a}-
\bm{M}_\mrm{b}^\dagger \right)^\dagger \right. \\
-\left( \bm{M}_\mrm{a}-\bm{M}_\mrm{b}^\dagger \right) 
\left. \cdot\left( \Im\{\bm{M}_\mrm{a}\} \right)^{-1} \cdot\frac{\delta\bm{M}_\mrm{a}}{2\iu} 
\cdot\left( \Im\{\bm{M}_\mrm{a}\} \right)^{-1}\right. \\ 
\left. \cdot\left( \bm{M}_\mrm{a}-\bm{M}_\mrm{b}^\dagger \right)^\dagger\right\}=\bm{0},
\end{multline}
and which can be reorganized as
\begin{multline}
\Re\left\{\left[\left( \bm{M}_\mrm{a}-\bm{M}_\mrm{b}^\dagger \right)\cdot\frac{\left( \Im\{\bm{M}_\mrm{a}\} \right)^{-1}}{2\iu} -\bm{I}\right] \right. \\
\left. \cdot\delta\bm{M}_\mrm{a}\cdot\left( \Im\{\bm{M}_\mrm{a}\} \right)^{-1}\cdot\left( \bm{M}_\mrm{a}-\bm{M}_\mrm{b}^\dagger \right)^\dagger
\right\}=\bm{0}.
\end{multline}
Assuming that both  $\Im\{\bm{M}_\mrm{a}\}>0$ and $\Im\{\bm{M}_\mrm{b}\}>0$ and hence that $\bm{M}_\mrm{a}\neq \bm{M}_\mrm{b}^\dagger$,
the optimal solution is obtained by requiring that the first line within parenthesis above vanishes, yielding
\begin{equation}
\Im\{\bm{M}_\mrm{a}\}=\frac{\bm{M}_\mrm{a}-\bm{M}_\mrm{b}^\dagger}{2\iu},
\end{equation}
or equivalently that $\bm{M}_\mrm{a}=\bm{M}_\mrm{b}$.
This solution implies that the minimum of $f(\bm{M}_\mrm{a})$ in \eqref{eq:fMadef} is given by
\begin{multline}\label{eq:minfMa}
f(\bm{M}_\mrm{a})\left|_{\bm{M}_\mrm{a}=\bm{M}_\mrm{b}}\right. = \int_{V} \bm{F}_\mrm{i}^*\cdot\left( \bm{M}_\mrm{b}-\bm{M}_\mrm{b}^\dagger \right)\\ 
\cdot\left( \Im\{\bm{M}_\mrm{b}\} \right)^{-1} \cdot\left( \bm{M}_\mrm{b}-\bm{M}_\mrm{b}^\dagger \right)^\dagger\cdot\bm{F}_\mrm{i}\diff v \\
= 4\int_{V}\bm{F}_\mrm{i}^*\cdot\Im\{\bm{M}_\mrm{b}\}\cdot\bm{F}_\mrm{i}\diff v.
\end{multline}
By substitution of the result \eqref{eq:minfMa} into the denominator of \eqref{eq:qdef}, it is finally seen that $q$ reaches its maximum at $1$, and hence $q\leq 1$.

\section{Spherical vector waves}\label{sect:spherical} 
\subsection{Definition of spherical vector waves}\label{sect:sphericaldef}
Consider a source-free homogeneous and isotropic medium with wave number $k=k_0\sqrt{\mu\epsilon}$. 
The electromagnetic field can then be expanded in spherical vector waves as
\begin{equation}\label{eq:EHsphdef}
\left\{\begin{array}{l}
\bm{E}(\bm{r})=\displaystyle\sum_{\tau,m,l}a_{\tau ml}{\bf v}_{\tau ml}(k\bm{r})+f_{\tau ml}{\bf u}_{\tau ml}(k\bm{r}), \vspace{0.2cm} \\
\bm{H}(\bm{r})=\displaystyle\frac{1}{\iu\eta_0\eta}\sum_{\tau,m,l}a_{\tau ml}{\bf v}_{\bar{\tau} ml}(k\bm{r})+f_{\tau ml}{\bf u}_{\bar{\tau} ml}(k\bm{r}),  
\end{array}\right.
\end{equation}
where ${\bf v}_{\tau ml}(k\bm{r})$ and ${\bf u}_{\tau ml}(k\bm{r})$ are the regular and the outgoing spherical vector waves, respectively, 
and $a_{\tau ml}$ and $f_{\tau ml}$ the corresponding multipole coefficients,
see \eg \cite{Newton1982,Bohren+Huffman1983,Bostrom+Kristensson+Strom1991,Arfken+Weber2001,Jackson1999,Kristensson2016}.
Here, $l=1,2,\ldots,$ is the multipole order, $m=-l,\ldots,l$, the azimuthal index and $\tau=1,2$, where
$\tau=1$ indicates a transverse electric (\textrm{TE}) magnetic multipole and $\tau=2$ a transverse magnetic (\textrm{TM}) electric multipole,
and $\bar{\tau}$ denotes the dual index, \ie $\bar{1}=2$ and $\bar{2}=1$.

The solenoidal (source-free) regular spherical vector waves are defined here by
\begin{multline}\label{eq:v1def}
\displaystyle{\bf v}_{1 ml}(k{\bm{r}})  =   \frac{1}{\sqrt{l(l+1)}}\nabla\times({\bm{r}}\mrm{j}_l(kr)\mrm{Y}_{ml}(\hat{\bm{r}})) \\
=   \mrm{j}_l(kr){\bf A}_{1 ml}(\hat{\bm{r}}),
\end{multline}
and 
\begin{multline}\label{eq:v2def}
{\bf v}_{2 ml}(k\bm{r})   =   \displaystyle \frac{1}{k}\nabla\times{\bf v}_{1 ml}(k\bm{r}) \\
 =\displaystyle\frac{(kr\mrm{j}_l(kr))^{\prime}}{kr}{\bf A}_{2 ml}(\hat{\bm{r}})+\sqrt{l(l+1)}\frac{\mrm{j}_l(kr)}{kr}{\bf A}_{3 ml}(\hat{\bm{r}}),
\end{multline}
where $\mrm{Y}_{ml}(\hat{\bm{r}})$ are the spherical harmonics, ${\bf A}_{\tau ml}(\hat{\bm{r}})$ the vector spherical harmonics and $\mrm{j}_l(x)$ the spherical Bessel functions of order $l$,
\cf \cite{Bostrom+Kristensson+Strom1991,Arfken+Weber2001,Jackson1999,Newton2002,Olver+etal2010,Kristensson2016}. 
Here, $(\cdot)^\prime$ denotes a differentiation with respect to the argument of the spherical Bessel function.
The outgoing (radiating) spherical vector waves ${\bf u}_{\tau ml}(k{\bm{r}})$ are obtained by replacing
the regular spherical Bessel functions $\mrm{j}_l(x)$ above with the spherical Hankel functions of the first kind, $\mrm{h}_l^{(1)}(x)$,  
see \cite{Bostrom+Kristensson+Strom1991,Jackson1999,Olver+etal2010,Kristensson2016}.
It can be shown that any one of the vector spherical waves ${\bf w}_{\tau ml}(k\bm{r})$ defined above satisfy the following curl properties
\begin{equation}\label{eq:w12cross}
\nabla\times {\bf w}_{\tau ml}(k\bm{r})=k{\bf w}_{\bar{\tau} ml}(k\bm{r}),
\end{equation}
and hence the source-free Maxwell's equations (vector Helmholtz equation) in free space, \ie 
\begin{equation}
\nabla\times\nabla\times {\bf w}_{\tau ml}(k\bm{r})=k^2 {\bf w}_{\tau ml}(k\bm{r}).
\end{equation}

The vector spherical harmonics ${\bf A}_{\tau ml}(\hat{\bm{r}})$ are given by
\begin{equation}\label{eq:Adef}
\left\{\begin{array}{lll}
{\bf A}_{1ml}(\hat{\bm{r}})  &  =   & \displaystyle\frac{1}{\sqrt{l(l+1)}}\nabla\times\left( \bm{r}\mrm{Y}_{ml}(\hat{\bm{r}}) \right) \\
     &    =  &  \displaystyle\frac{1}{\sqrt{l(l+1)}}\nabla\mrm{Y}_{ml}(\hat{\bm{r}})\times\bm{r},   \vspace{0.2cm}\\
{\bf A}_{2ml}(\hat{\bm{r}})  &    =  &   \hat{\bm{r}}\times{\bf A}_{1ml}(\hat{\bm{r}})  \vspace{0.2cm}\\
     &  =  &   \displaystyle\frac{1}{\sqrt{l(l+1)}}r\nabla\mrm{Y}_{ml}(\hat{\bm{r}}),  \vspace{0.2cm}\\
{\bf A}_{3ml}(\hat{\bm{r}})  &  =  &  \hat{\bm{r}}\mrm{Y}_{ml}(\hat{\bm{r}}), 
\end{array}\right.
\end{equation}
where $\tau=1,2,3$, and where the spherical harmonics $\mrm{Y}_{ml}(\hat{\bm{r}})$ are given by
\begin{equation}
\mrm{Y}_{ml}(\hat{\bm{r}})=\sqrt{\frac{2l+1}{4\pi}}\sqrt{\frac{(l-m)!}{(l+m)!}}\mrm{P}_{l}^m(\cos\theta)\eu^{{\rm i}m\phi}, 
\end{equation}
and where $\mrm{P}_{l}^m(x)$ are the associated Legendre functions \cite{Arfken+Weber2001,Jackson1999,Olver+etal2010}.
The vector spherical harmonics are orthonormal on the unit sphere, and hence
\begin{equation}\label{eq:Aorthonormal}
\int_{\Omega_0}{\bf A}_{\tau ml}^*(\hat{\bm{r}})\cdot{\bf A}_{\tau^\prime m^\prime l^\prime }(\hat{\bm{r}})\diff \Omega
=\delta_{\tau\tau^\prime}\delta_{mm^\prime}\delta_{ll^\prime},
\end{equation}
where $\Omega_0$ denotes the unit sphere and $\diff\Omega=\sin\theta\diff\theta\diff\phi$. 

\subsection{Lommel integrals for spherical Bessel functions}\label{sect:Lommel}
Let $\mrm{s}_l(kr)$ denote an arbitrary linear combination of spherical Bessel and Hankel functions.
Based on the two Lommel integrals for cylinder functions, \cf \cite[Eqs.~(10.22.4) and (10.22.5) on p.~241]{Olver+etal2010} and \cite[Eqs.~(8) and (10) on p.~134]{Watson1995},
the following indefinite Lommel integrals can be derived for spherical Bessel functions
\begin{equation}\label{eq:firstLommel}
\int\left|\mrm{s}_l(kr)\right|^2r^2\diff r
=r^2\frac{ \Im\!\left\{k\mrm{s}_{l+1}(kr) \mrm{s}_{l}^*(kr) \right\}}{\Im\!\left\{k^2\right\}},
\end{equation}
where $k$ is complex-valued ($k\neq k^*$), \cf \cite[Eq.~(A.15) on p.~11]{Nordebo+etal2017a}, and
\begin{equation}\label{eq:secondLommel}
\int\mrm{s}_l^2(kr)r^2\diff r 
=\frac{1}{2}r^3\left(\mrm{s}_l^2(kr)-\mrm{s}_{l-1}(kr)\mrm{s}_{l+1}(kr) \right),
\end{equation}
where $k$ is either real-valued or complex-valued.
Furthermore, by using the recursive relationships
\begin{equation}\label{eq:recursive}
\hspace{-0.05cm}\left\{\begin{array}{l}
\displaystyle \frac{\mrm{s}_l(kr)}{kr}=\frac{1}{2l+1}\left(\mrm{s}_{l-1}(kr)+\mrm{s}_{l+1}(kr) \right), \vspace{0.2cm} \\
\displaystyle \mrm{s}_l^\prime(kr)=\frac{1}{2l+1}\left(l\mrm{s}_{l-1}(kr)-(l+1)\mrm{s}_{l+1}(kr) \right), 
\end{array}\right.
\end{equation}
where $l=0,\pm 1,\pm 2,\ldots$, \cf \cite[Eqs. 10.1.19-20]{Abramowitz+Stegun1970}, 
it can be shown that
\begin{multline}\label{eq:recursiveLommel}
\int\left(\left|\frac{\mrm{s}_{l}(kr)}{kr}+\mrm{s}_{l}^\prime(kr)\right|^2 
 +l(l+1)  \left|\frac{\mrm{s}_{l}(kr)}{kr}\right|^2\right)r^2\diff r  \\
 =\frac{1}{2l+1}\int \left( (l+1)\left|\mrm{s}_{l-1}(kr)\right|^2+l\left|\mrm{s}_{l+1}(kr)\right|^2\right)r^2\diff r,
\end{multline}
for all values of $l=0,\pm 1,\pm 2,\ldots$.
It can be shown similarly that \eqref{eq:recursiveLommel} is valid with $|\cdot|^2$ replaced for $(\cdot)^2$.

\subsection{Orthogonality over a spherical volume}\label{sect:sphericalorthvol}
Due to the orthonormality of the vector spherical harmonics \eqref{eq:Aorthonormal}, it follows that
the regular spherical vector waves defined in \eqref{eq:v1def} and \eqref{eq:v2def} are orthogonal over a spherical volume $V_a$ of radius $a$, yielding
\begin{equation}\label{eq:vorthogonal}
\displaystyle\int_{V_a}{\bf v}_{\tau ml}^*(k{\bm{r}})\cdot{\bf v}_{\tau^\prime m^\prime l^\prime}(k{\bm{r}})\diff v 
=\displaystyle\delta_{\tau\tau^\prime}\delta_{mm^\prime}\delta_{ll^\prime}W_{\tau l}(k,a),
\end{equation}
where 
\begin{equation}\label{eq:Wtauldef}
W_{\tau l}(k,a)=\int_{V_a}\left|{\bf v}_{\tau ml}(k\bm{r})\right|^2\diff v,
\end{equation}
for $\tau=1,2$ and $l\geq 1$ and where $\diff v=r^2\diff\Omega\diff r$.
For complex-valued arguments $k\neq k^*$, $W_{1l}(k,a)$ is obtained from \eqref{eq:firstLommel} as
\begin{equation}\label{eq:W1ldef}
W_{1l}(k,a)=\int_{0}^{a}\left|\mrm{j}_{l}(kr)\right|^2 r^2\diff r 
=\frac{a^2 \Im\!\left\{k\mrm{j}_{l+1}(ka) \mrm{j}_{l}^*(ka) \right\}}{\Im\!\left\{k^2\right\}},
\end{equation}
and from \eqref{eq:recursiveLommel} follows that
\begin{multline}\label{eq:W2ldef}
W_{2 l}(k,a) \\
=\int_{0}^{a}\left(\left|\frac{\mrm{j}_{l}(kr)}{kr}+\mrm{j}_{l}^\prime(kr)\right|^2 
 +l(l+1)  \left|\frac{\mrm{j}_{l}(kr)}{kr}\right|^2\right)r^2\diff r  \\
 =\frac{1}{2l+1}\left[(l+1)W_{1,l-1}(k,a)+lW_{1,l+1}(k,a) \right].  
\end{multline}
For real-valued arguments ($k^*=k$), $W_{1l}(k,a)$ and $W_{2l}(k,a)$ can be calculated similarly by using  \eqref{eq:secondLommel} and \eqref{eq:W2ldef}, respectively.

\subsection{Orthogonality over a spherical surface}\label{sect:sphericalorthsurf}
Based on the properties of the spherical vector waves described in  Appendix \ref{sect:sphericaldef}, the
following orthogonality relationships regarding their cross products on a spherical surface can be derived as 
\begin{multline}\label{eq:orthcrosssph1}
\int_{\partial V_a}{\bf w}_{\tau ml}(k\bm{r})\times {\bf z}_{\bar{\tau} m^\prime l^\prime}^*(k\bm{r})\cdot\hat{\bm{r}}\diff S \\
=a^2\delta_{mm^\prime}\delta_{ll^\prime}
\left\{\begin{array}{ll}
 \displaystyle  w_l(ka)\left(\frac{(ka z_l(ka))^\prime}{ka} \right)^*  & \tau=1, \vspace{0.2cm} \\
\displaystyle  -\left(\frac{(ka w_l(ka))^\prime}{ka} \right)z_l^*(ka) & \tau=2,
\end{array}\right.
\end{multline}
and 
\begin{equation}\label{eq:orthcrosssph2}
\int_{\partial V_a}{\bf w}_{\tau ml}(k\bm{r})\times {\bf z}_{\tau m^\prime l^\prime}^*(k\bm{r})\cdot\hat{\bm{r}}\diff S=0,
\end{equation}
for $\tau=1,2$.
Here, $\partial V_a$ is the spherical surface of radius $a$,  $w_l(ka)$ and $z_l(ka)$ are either of $\mrm{j}_{l}(ka)$ or $\mrm{h}_{l}^{(1)}(ka)$, 
and ${\bf w}_{\tau ml}(k\bm{r})$ and ${\bf z}_{\tau ml}(k\bm{r})$ are the corresponding spherical vector waves, respectively.

\subsection{Mie theory}\label{sect:Mietheory}

Consider the scattering of the electromagnetic field due to a layered sphere consisting of $N$ layers made of isotropic materials. Let $a_i$, $\epsilon_{i}$, $\mu_i$, $k_i=k_0\sqrt{\mu_i\epsilon_i}$ and $\eta_i=\sqrt{\mu_i/\epsilon_i}$ (for $i=1,\ldots,N$) denote the radii, the relative permittivities, the relative permeabilities, the wave numbers, and the relative wave impedances of each of $N$ layers of the sphere, respectively. The scatterer of the total radius $a=a_N$ is embedded in the medium characterized by the wave number $k_\mrm{b}=k_0\sqrt{\mu_\mrm{b}\epsilon_\mrm{b}}$ and the relative wave impedance $\eta_\mrm{b}=\sqrt{\mu_\mrm{b}/\epsilon_\mrm{b}}$, where $\epsilon_\mrm{b}$ and $\mu_\mrm{b}$ denote the permittivity and the permeability of the surrounding medium, respectively. 

The electric and magnetic fields in each layer, for $a_{i-1}<r<a_{i}$, $i=1,\ldots,N$, can be expanded in spherical vector waves as
\begin{equation}
\left\{
\begin{array}{l}
\bm{E}(\bm{r}) = \displaystyle \sum_{\tau,m,l} A_{\tau ml}^{(i)}\left({\bf v}_{\tau ml}(k_i\bm{r})+t_{\tau l}^{(i-1)}{\bf u}_{\tau ml}(k_i\bm{r})   \right), \vspace{0.2cm} \\
\bm{H}(\bm{r}) = \displaystyle \frac{1}{\iu\eta_0\eta_i} \sum_{\tau,m,l} A_{\tau ml}^{(i)}\left({\bf v}_{\overline{\tau} ml}(k_i\bm{r})+t_{\tau l}^{(i-1)}{\bf u}_{\overline{\tau} ml}(k_i\bm{r})   \right),
\end{array}
\right.
\end{equation}
where $A_{\tau ml}^{(i)}$ and $t_{\tau l}^{(i)}$ are the corresponding multipole coefficients and transition matrices, respectively, see \eg \cite[Eq.~(8.16) on p.~436]{Kristensson2016}.

By matching the boundary conditions for the tangential electric and magnetic fields at each boundary interface $a_i$, the transition matrix for scattering in the corresponding layer $i=1,\ldots,N$ can be determined as
\begin{widetext}
\begin{equation}\label{eq:ttaulApp2} 
t_{\tau l}^{(i)} = -\frac{m_{\tau}^{(i)}\left(\psi_l(x_i)+t_{\tau l}^{(i-1)}\xi(x_i)	\right)\psi^{\prime}_l(y_i)-
\left(\psi_l^\prime(x_i)+t_{\tau l}^{(i-1)}\xi^\prime(x_i)	\right)\psi_l(y_i)}
{m_{\tau}^{(i)}\left(\psi_l(x_i)+t_{\tau l}^{(i-1)}\xi(x_i)	\right)\xi^{\prime}_l(y_i)-
\left(\psi_l^\prime(x_i)+t_{\tau l}^{(i-1)}\xi^\prime(x_i)	\right)\xi_l(y_i)},
\end{equation}
\end{widetext}
for $\tau=1,2$, where $m_1^{(i)}=\eta_i/\eta_{i+1}$ and $m_2^{(i)}=\eta_{i+1}/\eta_i$, $x_i=k_ia_i$ and $y_i=k_{i+1}a_i$ denote the electrical radius in terms of the material paratmers on the internal and external sides of the interface $a_i$, respectively, see \cite[p.~437]{Kristensson2016}.
Note that  $k_{N+1}=k_\mrm{b}$, $\eta_{N+1}=\eta_\mrm{b}$, and $t_{\tau l}^{(0)}=0$ due to the non-singular behavior of the electric field at the origin, \ie $\bm{r}=\bm{0}$. Hereby, the expression for the transition matrix \eqref{eq:ttaulApp2}  reduces to the result for an homogeneous isotropic sphere, see \cite[Eqs.~(4.52) on p.~100]{Bohren+Huffman1983} and \cite[Eq.~(8.7) on p.~420]{Kristensson2016}, and the multipole coefficients $A_{\tau ml}^{(1)}$ agree with the coefficients $a_{\tau ml}$ in \eqref{eq:EHsphdef}.

Let $\bm{E}_\mrm{i}(\bm r)=\bm{E}_0\eu^{\mrm{i} k\hat{\bm{k}}\cdot\bm{r}}$ describe a plane wave
with vector amplitude $\bm{E}_0$ and propagation direction $\hat{\bm{k}}$.
It can be shown that the corresponding multipole expansion coefficients are given by
\begin{equation}\label{eq:aiplanewave}
a_{\tau ml}^\mrm{i}=4\pi\iu^{l-\tau+1}\bm{E}_0\cdot{\bf A}_{\tau ml}^*(\hat{\bm{k}}),
\end{equation}
for $\tau=1,2$, $l=1,2,\ldots$, and $m=-l,\ldots,l$, and where the vector spherical harmonics ${\bf A}_{\tau ml}(\hat{\bm{k}})$
are defined as in \eqref{eq:Adef}, see also \cite[Eq.~(7.28) on p.~375]{Kristensson2016}. 
Based on the sum identities for the vector spherical harmonics \cite[Eqs.~(A17) and (A18)]{Nordebo+etal2019a},
it can be shown that
\begin{equation}\label{eq:sumoverm}
\sum_{m=-l}^l \left| a_{\tau ml}^\mrm{i} \right|^2=2\pi (2l+1)\left| \bm{E}_0\right|^2,
\end{equation}
for $\tau=1,2$.


\end{document}